\documentclass[twocolumn,showpacs,preprintnumbers,amsmath,amssymb]{revtex4-1}

\usepackage{graphicx}
\usepackage{dcolumn}
\usepackage{bm}
\usepackage{amsmath,amsthm,amssymb}   
\usepackage{mathrsfs}

\def\avg#1{\big <#1 \big >}
\def\lesssim{\ \raise.3ex\hbox{$<$}\kern-0.8em\lower.7ex\hbox{$\sim$}\ }
\def\gesim{\ \raise.3ex\hbox{$>$}\kern-0.8em\lower.7ex\hbox{$\sim$}\ }

\font\scripti=cmmi7
\font\scriptscripti=cmmi5
\def\sib#1{\setbox0 = \hbox{\scripti #1}
  \kern-.02em\copy0\kern-\wd0
  \kern.04em\box0} 
\def\ssib#1{\setbox0 = \hbox{\scriptscripti #1}
  \kern-.02em\copy0\kern-\wd0
  \kern.04em\box0} 
\font\tenib=cmmib10 
\skewchar\tenib='177 \skewchar\tenib='177 \skewchar\tenib='177
\def\pbold#1{\setbox0 = \hbox{$ #1 $}
  \kern-.022em\copy0\kern-\wd0
  \kern.011em\copy0\kern-\wd0
  \kern.011em\copy0\kern-\wd0
  \kern.011em\copy0\kern-\wd0
  \kern.011em\box0} 

\begin{document}
\widetext
\title{Dynamical instability of a driven-dissipative electron-hole condensate in the BCS-BEC-crossover region}
\author{Ryo Hanai}
\email{hanai@acty.phys.sci.osaka-u.ac.jp}
\affiliation{Department of Physics, Osaka University, Toyonaka 560-0043, Japan} 
\author{Peter B. Littlewood}
\affiliation{Physical Sciences and Engineering, Argonne National Laboratory, Argonne, Illinois 60439, USA} 
\affiliation{James Franck Institute and Department of Physics, University of Chicago, IL 60637, USA} 
\author{Yoji Ohashi}
\affiliation{Department of Physics, Keio University, 3-14-1 Hiyoshi, Kohoku-ku, Yokohama 223-8522, Japan} 
\date{\today}
\begin{abstract}
We present a stability analysis on a driven-dissipative electron-hole condensate in the BCS (Bardeen-Cooper-Schrieffer)-BEC (Bose-Einstein-condensation)-crossover region. Extending the combined BCS-Leggett theory with the generalized random phase approximation (GRPA) to the non-equilibrium case by employing the Keldysh formalism, we show that the pumping-and-decay of carriers causes a depairing effect on excitons. 
This phenomenon gives rise to an attractive interaction between excitons in the BEC regime, as well as a supercurrent that anomalously flows anti-parallel to $\nabla \theta(\bm r)$ (where $\theta(\bm r)$ is the phase of the condensate) in the BCS regime, both leading to dynamical instabilities of an exciton-BEC. 
Our result suggests that substantial region of the exciton-BEC phase in the phase diagram (in terms of the interaction strength and the decay rate) is unstable.  
\end{abstract}
\maketitle

\section{Introduction}

A gas mixture of electrons and holes in a highly excited semiconductor provides a useful playground to study various fundamental many-body phenomena, such as the exciton-Mott transition (crossover)\cite{Mott1961,Zimmermann1978,Kwong2009,Yoshioka2012,Fehrenbach1982,Suzuki2012,Sekiguchi2015}, as well as the electron-hole liquid and droplet formation\cite{Combescot1972,Brinkman1973,Combescot1976,Jeffries}. Perhaps the most striking phenomenon is the exciton-BEC\cite{Blatt1962,Keldysh1968}. 
Indeed, signs of this phase have been reported by some experimental groups, such as a sudden enhancement of two-body inelastic scattering\cite{Yoshioka2011}, and the appearance of an anomaly in its spatial distribution\cite{Stolz2012} at low temperatures $(\sim O(100 {\rm mK}))$\cite{Stolz2012,Yoshioka2013}.
Thus, although further analyses are still necessary for these results, the achievement of exciton-BEC is very promising.

The realization of an exciton-BEC would enable us to examine the so-called BCS-BEC crossover\cite{Eagles1969,Leggett1980}, where the character of an exciton condensate continuously changes from the weak-coupling BCS-type (where electron-hole pairs are largely overlapping one another as in ordinary superconductors) to BEC of tightly bound excitons, with decreasing the carrier density $n$\cite{Comte1982,Nozieres1982,Nozieres1985}. Although a similar crossover phenomenon has already been discussed in cold Fermi gas physics\cite{Regal2004,Zwierlein2004}, a crucial difference from this atomic case is that the exciton case is essentially in the non-equilibrium state\cite{MoskalenkoSnoke}, because continuous pumping to compensate carrier loss is always necessary to sustain the system. Thus, the exciton system would be useful for the study of interplay between strong correlations and non-equilibrium effects in the BCS-BEC-crossover region. 
Since non-equilibrium Fermi condensates have been discussed in various fields, such as an exciton-polariton condensate\cite{Kasprzak2006}, quench dynamics of an ultracold gas\cite{Yuzbashyan2015}, as well as neutron-star cooling\cite{Yakovlev2005}, the realization of an exciton-BEC would also contribute to study of these systems as well.

A crucial non-equilibrium effect is the instability of a condensate. One well-known example is the Landau instability, where the supercurrent state becomes unstable when the flow velocity exceeds a critical value. Thus, it will not be too surprising for such instability to occur in a driven-dissipative system as well. Indeed, a dynamical instability has recently been observed in a polariton gas\cite{Bobrovska,Baboux2017}, which has also been theoretically studied in a boson model\cite{Wouters2007}, as well as a Dicke model\cite{Szymanska2006}, consisting of bosons coupled to localized fermions by disorder. However, no studies have been done on how strong correlations in the Fermi degrees of freedom affect the stability of a non-equilibrium condensate.

In this paper, by employing the combined BCS-Leggett theory\cite{Leggett1980} with the generalized random phase approximation\cite{Anderson1958,Schrieffer} extended to the Keldysh formalism\cite{Rammer}, we present two mechanisms of dynamical instabilities that occur in this system.
One is triggered by an \textit{attractive} interaction between excitons, appearing in the BEC regime, and the other arises in the BCS regime due to an anomalous supercurrent that flows \textit{anti}-parallel to the twist of the phase of the condensate $\nabla \theta(\bm r)$.
We show that both the instabilities originate from non-equilibrium-induced pair-breaking effects\cite{Hanai2016}.
Substantial region of the phase diagram, in terms of the interaction strength and the decay rate, is found to be unstable.

\begin{figure*}
\begin{center}
\includegraphics[width=0.6\linewidth,keepaspectratio]{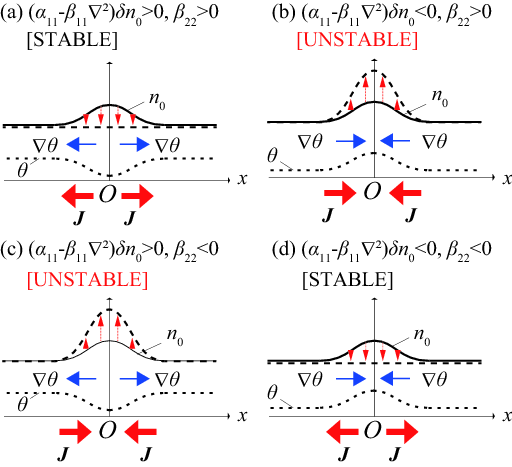}
\end{center}
\caption{(Color online) Schematic explanation of the hydrodynamics of the condensate fraction $n_0(\bm r,t)$ (solid line) and the phase $\theta(\bm r,t)$ of the condensate (dotted line).
Following Eq. (\ref{tGL}), the local enhancemet of the condensate fraction induces a spatial modulation of the phase $\theta(\bm r,t)$.
Then, following Eq. (\ref{continuity}), the phase gradient $\nabla\theta$ drives a supercurrent $\bm J\propto \beta_{22}\nabla\theta$. 
As described by the dashed lines, this supercurrent either flows away from the enhanced spot to damp the condensate fraction back to the steady-state (as in cases (a) and (d)), or amplify the local fluctuation even more (as in cases (b) and (c)), destabilizing the steady-state.
}
\label{fig_instability}
\end{figure*}

Figure \ref{fig_instability} schematically illustrates the hydrodynamics of the dynamical instabilities we found in this paper, explaining how fluctuations of the Bose-condensate fraction $\delta n_0(\bm r,t)$ from its steady-state value $n_0$ (as well as the phase fluctuations of the condensate $\delta\theta(\bm r,t)$) grow as a function of time in the dynamically unstable regimes.
From our microscopic analysis, the driven-dissipative Bose-condensate are found to approximately obey the hydrodynamic equations (which we derive in Sec. \ref{sec_result}),
\begin{eqnarray}
\frac{1}{\eta_{12}}(\alpha_{11}-\beta_{11}\nabla^2)\delta n_0(\bm r,t) 
+2 n_0 \partial_t \delta\theta(\bm r,t)
&=&0, 
\label{tGL} 
\\
\partial_t\delta n_0(\bm r,t)+\nabla\cdot\bm J(\bm r,t) 
&=&0,
\label{continuity}
\end{eqnarray}
where Eq. (\ref{continuity}) can be regarded as a continuity equation associated with a ``supercurrent''\cite{NOTEcontinuity}
\begin{eqnarray}
\bm J(\bm r,t) = 2n_0\frac{\beta_{22}}{\eta_{12}}\nabla \delta\theta(\bm r,t).
\label{supercurrent}
\end{eqnarray}
The (real) coefficients $\alpha_{ij}, \beta_{ij}$, and $\eta_{ij}$ are derived from our microscopic calculations we present in this paper.
These equations (\ref{tGL}), (\ref{continuity}) essentially have the same structure as the time-dependent Gross-Pitaevskii (GP) equations\cite{Varma2002},
\begin{eqnarray}
\bigg[U_{\rm B}n_0-\frac{\nabla^2}{2m_{\rm B}}
\bigg]
\delta n_0(\bm r,t) 
+2 n_0 \partial_t \delta\theta(\bm r,t)
&=&0, 
\label{tGLGP} 
\\
\partial_t\delta n_0(\bm r,t)+\nabla \cdot \bm J_{\rm B}(\bm r,t) 
&=&0,
\label{continuityGP}
\end{eqnarray}
for a repulsively interacting BEC\cite{PethickSmith}. 
Here, $U_{\rm B}$ is a repulsive interaction between bosons, $m_{\rm B}$ is the boson mass, and 
\begin{eqnarray}
\bm J_{\rm B}(\bm r,t)=\frac{n_0}{m_{\rm B}}\nabla\delta\theta(\bm r,t),
\end{eqnarray}
is a supercurrent of the bosons.

In the conventional case where all coefficients $\alpha_{ij},\beta_{ij}$, and $\eta_{ij}$ are positive, the uniform steady-state is stable against perturbations (Fig. \ref{fig_instability}(a)). 
In this case, the supercurrent $\bm J$ 
flows away from the locally enhanced condensate fraction $\delta n_0(\bm r,t)$ to stabilize the system back to the steady-state.
However, we show in this paper both numerically and analytically that the non-equilibrium nature of the electron-hole Bose-condensate can make coefficients such as $\alpha_{11}$ and $\beta_{22}$ switch to negative to give rise to anamolous hydrodynamics. 
Negative $\alpha_{11}$ correponds to the arise of an effective \textit{attractive} interaction between the excitons (Compare our hydrodynamic equation (\ref{tGL}) and the GP equation (\ref{tGLGP}).), and negative $\beta_{22}$ corresponds to the arise of an anomalous supercurrent that flows \textit{anti}-parallel to the phase gradient $\nabla\theta$ (Eq. (\ref{supercurrent})).
As illustrated in Figs. \ref{fig_instability}(b) and (c), these can give rise to a flow of supercurrent that amplifies the fluctuations, resulting in dynamical instabilities.

The rest of our paper is organized as follows. In Sec. \ref{sec_formalism}, we explain our driven-dissipative electron-hole model, as well as our formalism.
In Sec. \ref{sec_result}, we examine how the non-equilibrium nature of this system affect the Bose-condensed phase. 
We show how non-equilibrium induced pair-dissociation lead to dynamical instabilities, in a wide range of parameter region.
In Sec. \ref{sec_summary}, we give the concluding remarks and raise some important remaining problems.
Throughout our paper, we set $\hbar=k_{\rm B}=1$.

\section{Model and formalism}\label{sec_formalism}

\begin{figure}
\begin{center}
\includegraphics[width=0.75\linewidth,keepaspectratio]{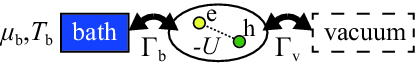}
\end{center}
\caption{(Color online) Model driven-dissipative electron-hole gas. An electron-hole system with an attractive interaction $-U$ is coupled to a pumping bath with the transfer matrix element $\Gamma_{\rm b}$, as well as a vacuum with the transfer matrix element $\Gamma_{\rm v}$. The pumping bath is assumed to be in a equilibrium state at the temperature $T_{\rm b}$, where the particle distribution obeys the Fermi distribution function $f_{\rm b}(\omega)=1/[e^{(\omega-\mu_{\rm b})/T_{\rm b}}+1]$, where $\mu_{\rm b}$ is the chemical potential. The particle distribution $f_{\rm v}(\omega)$ in the vacuum vanishes.}
\label{fig_model}
\end{figure}

The model driven-dissipative electron-hole gas we consider is illustrated in Fig. \ref{fig_model}. The corresponding Hamiltonian is given by, $H=H_{\rm s}+H_{\rm t}+H_{\rm env}$, where  
\begin{eqnarray}
H_{\rm s}&=&\sum_{\bm p} \Psi^\dagger_{\bm p} [\varepsilon_{\bm p} \tau_3 - \Delta(t) \tau_+ - \Delta^*(t) \tau_-]\Psi_{\bm p}
\nonumber\\
&-&
U\sum_{\bm q}\rho_{\bm q}^+\rho_{-{\bm q}}^{-},
\label{Hs}
\end{eqnarray}
describes an electron-hole gas in the exciton-BEC phase.
$\Psi^\dagger_{\bm p}=(a^\dagger_{\bm p,{\rm e}},a_{-\bm p,{\rm h}})$ is a Nambu field, consisting of an electron creation ($a^\dagger_{\bm p,{\rm e}}$) and a hole annihilation operator ($a_{-\bm p,{\rm h}}$). Electrons and holes are assumed to have the same mass $m$ and kinetic energy $\varepsilon_{\bm p}={\bm p}^2/(2m)$. 
Pauli matrices $\tau_s$ ($s=1,2,3$) and $\tau_{\pm}=[\tau_1 \pm i\tau_2]/2$ act on the Nambu-space, and $\rho_{\bm q}^s=\sum_{\bm p}\Psi^\dagger_{\bm p+\bm q/2}\tau_s \Psi_{\bm p-\bm q/2}$ is a generalized density operator.

In this paper, we assume an attractive contact-type interaction $-U~(<0)$ between electrons and holes for simplicity.
Although Eq. (\ref{Hs}) ignores long-range Coulomb interaction, since the gapless Goldstone mode appears in the exciton-BEC phase regardless of whether the interaction is short- or long-ranged\cite{Cote1988} (Note that since electron-hole \textit{pairs} are neutral, the Anderson-Higgs mechanism is absent.), we expect that our model can still capture low-energy properties of the exciton-BEC, at least qualitatively.
In this model, exciton-BEC is characterized by the order parameter $\Delta(t)=U\sum_{\bm p}\avg{a_{-\bm p,{\rm h}}(t)a_{\bm p,{\rm e}}(t)}$. 
Following the conventional prescription\cite{Randeria}, we measure the interaction strength in terms of the scattering length $a_s$, given by $4\pi a_s/m=-U/[1-U\sum_{\bm p}^{p_{\rm c}}1/(2\varepsilon_{\bm p})]$ (where $p_{\rm c}$ is a momentum cutoff). 
In this scale, the weak-coupling BCS regime and the strong-coupling BEC regime are characterized as $(k_{\rm F}a_s)^{-1}\lesssim 0$ and $(k_{\rm F}a_s)^{-1}\gesim 0$, respectively (where $k_{\rm F}=(3\pi^2N)^{1/3}$ is the Fermi momentum with $N$ being the particle number in the system).

Incoherent pumping and decay of the carriers are, respectively, driven by a coupling $\Gamma_{\rm b}$ to a free Fermi bath and a coupling $\Gamma_{\rm v}$ to the vacuum as done in Refs.\cite{Szymanska2006,Hanai2016,Yamaguchi2012}. 
These tunneling processes are described by, 
\begin{eqnarray}
H_{\rm t} = \sum_{\lambda={\rm b,v}}\sum_{{\bm p},{\bm q}}\sum_i 
[\Gamma_\lambda \Phi^{\lambda\dagger}_{\bm q} \tau_3 \Psi_{\bm p}e^{-i{\bm q}\cdot{\bm R}_i}e^{i{\bm p}\cdot{\bm r}_i} + {\rm h.c.}].
\label{Ht}
\end{eqnarray}
Here, a particle in the exciton BEC system at ${\bm r}_i$ is assumed to tunnel to the bath or vacuum at ${\bm R}_i$. The bath and vacuum are described by
\begin{eqnarray}
H_{\rm env}=
\sum_{\lambda={\rm b,v}}\sum_{\bm q}\Phi_{\bm q}^{\lambda\dagger}
\varepsilon_{\bm q}^{\lambda} \tau_3 \Phi_{\bm q}^{\lambda}.
\label{Henv}
\end{eqnarray}
The Nambu field $\Phi_{\bm q}^{\rm b(v)} = (c^{\rm b(v)}_{\bm q,{\rm e}}, c^{{\rm b(v)}\dagger}_{-\bm q,{\rm h}} )^{\rm T}$ consists of an electron annihilation operator $c^{\rm b(v)}_{\bm q,{\rm e}}$ and a hole creation operator $c^{\rm b(v)\dagger}_{-\bm q,{\rm h}}$ in the bath (vacuum). $\varepsilon_{\bm q}^{\rm b(v)}$ is the kinetic energy of the bath (vacuum). 
We assume that the bath and vacuum are huge compared to the exciton-BEC system, and the former two are always in the equilibrium state. Carriers in the bath obey the Fermi distribution function $f_{\rm b}(\omega)=[e^{(\omega-\mu_{\rm b})/T_{\rm b}}+1]^{-1}$ (where $\mu_{\rm b},T_{\rm b}$ is the chemical potential and the temperature of the bath, respectively). 
In this paper, we only treat the $T_{\rm b}=0$ case. 
Since the vacuum has no particle, we assume a vanishing distribution $f_{\rm v}(\omega)= 0$. 

Our treatment of the pumping of excitons as a coupling to a free Fermi bath is, at a glance, quite different from an experimental situation, which is rather a photon pumping and its thermalization via phonon emission to the bulk semiconductor or exciton-exciton scattering.  
However, since the attachment of the bath to the system phenomenologically describes the injection and thermalization of carriers, we expect our model to capture the fundamental elements of the above processes.
Similarly, we expect our modelling of the decay process of excitons as a coupling to a vacuum (which in reality would be photon recombination) to capture the essence of the decay process, since the phase-breaking effect due to the loss of electron-hole pairs\cite{Szymanska2003} are taken into account.

Non-equilibrium effects on a strongly-interacting Bose-condensate are conveniently examined by using the single-particle Nambu-Keldysh Green's function $\hat {\mathscr G}$, obeying the Dyson's equation\cite{Rammer},
\begin{eqnarray}
\hat {\mathscr G}(\bm r,t;\bm r',t')
&=&
\hat G^{0}(\bm r-\bm r',t-t')
\nonumber\\
&+&
\int d\bm r_1 d\bm r_2 dt_1 dt_2 \hat G^{0}(\bm r-\bm r_1,t-t_1)
\nonumber\\
&\times&
\hat\Sigma(\bm r_1,t_1;\bm r_2,t_2)\hat {\mathscr G}(\bm r_2,t_2;\bm r',t').
\label{Dyson}
\end{eqnarray}
Here, $\hat G^0$ is a free Green's function and $\hat\Sigma$ is the self-energy that incorporates non-equilibrium and interaction effects. 
In our RAK representation\cite{Rammer}, the single-particle Green's function $\hat {\mathscr G}$, the self-energy $\hat \Sigma$, as well as a free Green's function $\hat G^0$, are represented as, 
\begin{eqnarray}
\hat {\mathscr G}
&=&
\left(
\begin{array}{cc}
{\mathscr G}_{aa} &{\mathscr G}_{ab} \\
{\mathscr G}_{ba} & {\mathscr G}_{bb}
\end{array}
\right)
=
\left(
\begin{array}{cc}
{\mathscr G}^{\rm R} & {\mathscr G}^{\rm K} \\
0                            & {\mathscr G}^{\rm A}
\end{array}
\right),\\
\hat \Sigma
&=&
\left(
\begin{array}{cc}
\Sigma_{aa} & \Sigma_{ab} \\
\Sigma_{ba} & \Sigma_{bb}
\end{array}
\right)
=
\left(
\begin{array}{cc}
\Sigma^{\rm R} & \Sigma^{\rm K} \\
0                     & \Sigma^{\rm A}
\end{array}
\right),
\end{eqnarray}
and
\begin{eqnarray}
&&
\hat G^0(\bm p,\omega)
=
\left(
\begin{array}{cc}
G^0_{aa} & G^0_{ab} \\
G^0_{ba} & G^0_{bb}
\end{array}
\right)
(\bm p,\omega)=
\left(
\begin{array}{cc}
G^{0{\rm R}} & G^{0{\rm K}} \\
0               & G^{0{\rm A}}
\end{array}
\right)
(\bm p,\omega)
\nonumber\\
&&=
\left(
\begin{array}{cc}
[\omega+i\delta -\varepsilon_{\bm p}\tau_3]^{-1} & -\pi i \tau_3 (1-2f(\omega\tau_3)) \delta(\omega-\varepsilon_{\bm p}\tau_3) \\
0                  & [\omega-i\delta -\varepsilon_{\bm p}\tau_3]^{-1}
\end{array}
\right),
\nonumber\\
\label{G0}
\end{eqnarray}
composed of the retarded, advanced, and the Keldysh component represented by the supersubscript ${\rm R}, {\rm A}$, and ${\rm K}$, respectively.
Here, $(\alpha,\alpha')=(a,a),(b,b),(a,b)$-component represents the retarded, advanced, and the Keldysh component, respectively, and $(b,a)$-component is zero.
While the retarded (advanced) component in the single-particle Green's function ${\mathscr G}^{\rm R}(=[{\mathscr G}^{\rm A}]^\dagger)$ gives information on the single-particle density of states of the system, the Keldysh component ${\mathscr G}^{\rm K}$ gives information on the occupancy of the density of states.
For details of the Keldysh Green's function method, see e.g., Ref.\cite{Rammer}.
$f(\omega)$ in Eq. (\ref{G0}) is the initial distribution of the system, which however, will be shown later to give no effects on the steady-state properties. 
Note that $G^0_{\alpha,\alpha'},{\mathscr G}_{\alpha,\alpha'}$ and $\Sigma_{\alpha,\alpha'}$ are matrices in Nambu space.
 
In order to proceed, an approximated treatment of the self-energy $\hat\Sigma$ is needed.
In this regard, we recall that, in the conventional equilibrium case, it is well-known that the so-called BCS-Leggett theory\cite{Leggett1980} well describes the BCS-BEC crossover physics in the ground state, at least qualitatively\cite{Horikoshi2017,Tajima2017}.
In this framework, unlike in the conventional BCS theory where the chemical potential is fixed at the Fermi energy $\mu=\varepsilon_{\rm F}$ (implicitly assuming a weak attractive interaction), 
the chemical potential of the system $\mu$ and the order parameter $\Delta_0$ are determined self-consistently by solving the gap equation 
\begin{eqnarray}
\frac{1}{U}
=\sum_{\bm p}\frac{1-2 n_{\rm F}(E_{\bm p})}{2E_{\bm p}},
\label{GAPeq}
\end{eqnarray}
together with the number equation,
\begin{eqnarray}
N&=&\sum_{\bm p}
\bigg[
\frac{1}{2}\Big(
1-\frac{\xi_{\bm p}}{E_{\bm p}}
\Big)
(1-n_{\rm F}(E_{\bm p}))
\nonumber\\
&& \ \ \ +
\frac{1}{2}\Big(
1+\frac{\xi_{\bm p}}{E_{\bm p}}
\Big)
n_{\rm F}(E_{\bm p})
\bigg].
\label{NUMeq}
\end{eqnarray}
Here, $n_{\rm F}(\omega)=[e^{\omega/T}+1]^{-1}$ is the Fermi distribution function, $E_{\bm p}=\sqrt{\xi_{\bm p}^2+\Delta_0^2}$ describes single-particle excitations, and $\xi_{\bm p}=\varepsilon_{\bm p}-\mu$.
These equations (\ref{GAPeq}) and (\ref{NUMeq}) are known to yield the two limiting cases; 
the weak-coupling limit ($(k_{\rm F}a_s)^{-1}\rightarrow-\infty$) where the conventional BCS results ($\Delta_0=(8/e^2)\varepsilon_{\rm F}e^{\pi/(2k_{\rm F}a_s)^{-1}}, \mu=\varepsilon_{\rm F}$) are obtained, 
and the strong-coupling limit ($(k_{\rm F}a_s)^{-1}\rightarrow +\infty$) where $\mu=-E_{\rm b}/2$ ($E_{\rm b}=-1/(ma_s^2)$ is the binding energy of an exciton) and $\Delta_0=\sqrt{16/(3\pi)}|\mu|^{1/4}\varepsilon_F^{3/4}$.
The strong-coupling limit (the so-called BEC limit) solution $\mu=-E_{\rm b}/2$ indicates that the system is well described by a BEC of strongly bound excitons, where the system earns half the binding energy when an electron or a hole is added to the system by forming an exciton.
These two limits continuously connects to each other by gradually increasing the interaction strength from the weak-coupling to the strong-coupling regime, describing the essential features of the BCS-BEC crossover physics\cite{Leggett1980}.

\par
\begin{figure}
\begin{center}
\includegraphics[width=0.9\linewidth,keepaspectratio]{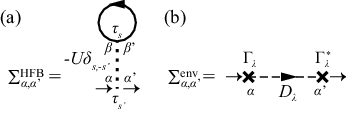}
\end{center}
\caption{(Color online) (a) The Hartree-Fock-Bogoliubov (HFB) self-energy $\hat\Sigma^{\rm HFB}$ and (b) the self-energy correction within the second Born approximation $\hat\Sigma^{\rm env}$. The solid and the dashed line are the single-particle Nambu-Keldysh Green's function of the system $\hat {\mathscr G}$ and the reservoirs $\hat D_{\lambda={\rm b,v}}$, respectively. 
The dotted line represents the coupling constant $-U$, and the cross represents the hopping between the system and the resevoirs $\Gamma_{\lambda={\rm b,v}}$.}
\label{figdiagram}
\end{figure}
\par

In the Nambu-Keldysh Green's function method, Eqs. (\ref{GAPeq}) and (\ref{NUMeq}) can be shown to be equivalent to approximating the self-energy within the Hartree-Fock-Bogoliubov (HFB) level, as given diagramatically in Fig. \ref{figdiagram}(a)\cite{Hanai2016,Schrieffer}.
We provide its explicit form in Appendix \ref{AppBCSLeggett}, and give here only its final form, 
\begin{eqnarray}
&&[\Sigma^{\rm HFB}(\bm r,t;\bm r',t')]^{\rm R}
=
iU\sum_{\bm p'}\int_{-\infty}^{\infty} \frac{d\omega'}{2\pi}
\nonumber\\
&&\times
\frac{1}{2}
[{\mathscr G}^{\rm K}_{12}(\bm r,t;\bm r,t+0^+)\tau_+
+{\mathscr G}^{\rm K}_{21}(\bm r,t;\bm r,t+0^+)\tau_-],
\label{SigHFBR}
\\
&&{}[\Sigma^{\rm HFB}(\bm r,t;\bm r',t')]^{\rm K}
=0,
\label{SigHFBK}
\\
&&{}[\Sigma^{\rm HFB}(\bm r,t;\bm r',t')]^{\rm A}
=
\big[
[\Sigma^{\rm HFB}(\bm r,t;\bm r',t')]^{\rm R}
\big]^\dagger.
\end{eqnarray}

To extend this framework to the driven-dissipative case, we further add the diagram shown in Fig. \ref{figdiagram}(b) as a self-energy that describes the tunneling effects to the bath and the vacuum within the second-order Born approximation\cite{Hanai2016}.
As derived in Appendix \ref{AppBCSLeggett}, this is given by,
\begin{eqnarray}
[\Sigma^{\rm env}(\omega)]^{\rm R}
&=&
- i\sum_{\lambda={\rm b,v}}\gamma_\lambda,
\label{SigenvR}
\\
{}[\Sigma^{\rm env}(\omega)]^{\rm K}
&=&
-2i\tau_3\sum_{\lambda={\rm b,v}}\gamma_\lambda[1-2f_\lambda(\omega\tau_3)],
\label{SigenvK}
\\
{}[\Sigma^{\rm env}(\omega)]^{\rm A}
&=&\big[
[\Sigma^{\rm env}(\omega)]^{\rm R}
\big]^\dagger,
\label{SigenvA}
\end{eqnarray}
where we have Fourier transformed the relative time $t-t'$, and have used the assumption that the bath and vacuum are large enough to stay in the equilibrium state during the dynamics.
Here, $\gamma_{\rm b(v)}=\pi N_{\rm t}\rho_{\rm b(v)}|\Gamma_{\rm b(v)}|^2$ characterizes the thermalization (decay) rate from the bath (vacuum). 
We have assumed a Markovian bath (vacuum), i.e., the density of states in the bath (vacuum) $\rho_{\rm b(v)}$ is constant. $N_{\rm t}$ is the number of tunneling paths.
The imaginary part in the retarded component (Eq. (\ref{SigenvR})) gives rise to a linewidth $\gamma=\gamma_{\rm b}+\gamma_{\rm v}$ to the single-particle excitation spectrum, and the Keldysh component (Eq. (\ref{SigenvK})) describes how the bath and vacuum distribution $f_\lambda(\omega)$ affects the distribution of the carriers in the system.

We now consider a uniform steady-state solution of the framework presented above, by imposing the steady-state ansatz\cite{Szymanska2006,Hanai2016,Yamaguchi2012},
\begin{eqnarray}
\Delta(t)=\Delta_0 e^{-2i\mu t}.
\label{Delta}
\end{eqnarray}
Without loss of generality, $\Delta_0>0$ is assumed to be real.
We can formally eliminate the time dependence of the order parameter in Eq. (\ref{Delta}) by employing the gauge transformation $(a_{\bm p,{\sigma={\rm e,h}}},c^\lambda_{\bm q,{\sigma={\rm e,h}}})=(\tilde a_{\bm p,{\sigma={\rm e,h}}},\tilde c^\lambda_{\bm q,{\sigma={\rm e,h}}})e^{-i\mu t}$. 
Practically, this is done by replacing $\varepsilon_{\bm p},\varepsilon_{\bm q}^\lambda$, and $\mu_{\rm b}$ by $\xi_{\bm p}=\varepsilon_{\bm p} - \mu, \xi_{\bm q}^\lambda=\varepsilon^\lambda_{\bm q}-\mu$, and $\mu_{\rm b}-\mu$, respectively\cite{Szymanska2006}.
After employing this gauge transformation, the uniform steady-state single-particle Green's function $\hat G(\bm r,t;\bm r',t')$ and the self-energy $\hat\Sigma_0(\bm r,t;\bm r',t')$ depend only on relative coordinates, i.e., $\bm r-\bm r'$ and $t-t'$, which simplifies the Dyson's equation (\ref{Dyson}) to
\begin{eqnarray}
\hat G(\bm p,\omega)
=\hat G^0(\bm p,\omega)+\hat G^0(\bm p,\omega)\hat\Sigma_0(\bm p,\omega)\hat G(\bm p,\omega).
\label{DysonNESS}
\end{eqnarray}

\begin{widetext}
The steady-state self-energy $\hat\Sigma_0$ is derived from Eqs. (\ref{SigHFBR})-(\ref{SigenvA}),
\begin{eqnarray}
\Sigma^{\rm R}_0(\bm p,\omega)
&=&
iU\sum_{\bm p'}\int \frac{d\omega'}{2\pi}
\frac{1}{2}
[
G^{\rm K}_{12}(\bm p',\omega') \tau_+
+G^{\rm K}_{21}(\bm p',\omega') \tau_-
]
- i\sum_{\lambda={\rm b,v}}\gamma_\lambda,
\label{SigR0}
=
-\Delta_0\tau_1 
- i\sum_{\lambda={\rm b,v}}\gamma_\lambda,
\label{SigR0Del}
\\
\Sigma^{\rm K}_0(\bm p,\omega)
&=&-2i\tau_3\sum_{\lambda={\rm b,v}}\gamma_\lambda[1-2f_\lambda(\omega\tau_3)],
\\
{}\Sigma^{\rm A}_0(\bm p,\omega)&=&[\Sigma^{\rm R}_0(\bm p,\omega)]^\dagger.
\label{SigA0}
\end{eqnarray}
In deriving Eq. (\ref{SigR0Del}), we have used the definition of the order parameter,
\begin{eqnarray}
\Delta_0 &=& U\sum_{\bm p}\avg{\tilde a_{-\bm p,{\rm h}} \tilde a_{\bm p,{\rm e}}}
=
-iU\sum_{\bm p}\frac{i}{2}
\big[
\avg{\tilde a_{-\bm p,{\rm h}}\tilde a_{\bm p,{\rm e}}}
-\avg{\tilde a_{\bm p,{\rm e}}\tilde a_{-\bm p,{\rm h}}}
\big]
=
 -iU\sum_{\bm p}\int\frac{d\omega'}{2\pi}
\frac{1}{2}
G^{\rm K}_{12}(\bm p,\omega'),
\end{eqnarray}
and the fact that $\Delta_0$ is assumed to be real.
From the Dyson's equation (\ref{DysonNESS}), the steady-state single-particle Green's function $\hat G(\bm p,\omega)$ is obtained as,
\begin{eqnarray}
&&
G^{\rm R}(\bm p,\omega)
=
[(\omega + i\delta)-\xi_{\bm p}\tau_3 -\Sigma_0^{\rm R}(\bm p,\omega)]^{-1}
\nonumber\\
&&=[(\omega + i\gamma) \bm 1 - \xi_{\bm p}\tau_3+\Delta_0 \tau_1]^{-1}
=
\frac{(\omega + i\gamma)\bm 1+\xi_{\bm p}\tau_3-\Delta_0\tau_1}{(\omega+ i\gamma)^2+E_{\bm p}^2}
=
\frac{1}{(\omega + i\gamma)^2+E_{\bm p}^2}
\left(
\begin{array}{cc}
\omega + i\gamma +\xi_{\bm p} & -\Delta_0 \\
-\Delta_0                                 & \omega + i\gamma -\xi_{\bm p}
\end{array}
\right),
\label{Gra}
\\
&&G^{\rm K}(\bm p,\omega)
=
G^{\rm R}(\bm p,\omega)\Sigma^{\rm K}_0(\omega)G^{\rm A}(\bm p,\omega)
+
(1+G^{\rm R}(\bm p,\omega)\Sigma^{\rm R}_0(\bm p,\omega))G^{0{\rm K}}(\bm p,\omega)
(1+\Sigma^{\rm A}_0(\bm p,\omega)G^{\rm A}(\bm p,\omega))
\label{Gk_1}
\\
&&=
\frac{-2i}{[(\omega-E_{\bm p})^2+\gamma^2][(\omega+E_{\bm p})^2+\gamma^2]}
\left(
\begin{array}{cc}
F(\omega)[(\omega+\xi_{\bm p})^2+\gamma^2]-F(-\omega)\Delta_0^2 
& -2\Delta_0[F_-(\omega)\omega +F_+ (\omega)(\xi_{\bm p}+i\gamma)]\\
-2\Delta_0[F_-(\omega)\omega +F_+ (\omega)(\xi_{\bm p}-i\gamma)]    
& F(\omega)\Delta_0^2-F(-\omega)[(\omega-\xi_{\bm p})^2+\gamma^2]
\end{array}
\right),
\nonumber\\
\label{Gk}
\\
&&
G^{\rm A}(\bm p,\omega) 
= \big[
G^{\rm R}(\bm p,\omega)
\big]^\dagger.
\end{eqnarray}
Here, $F_{\pm}(\omega)=[F(\omega)\pm F(-\omega)]/2$ 
with $F(\omega)=\sum_{\lambda={\rm b},{\rm v}}\gamma_\lambda[1-2f_\lambda(\omega)]$.
\end{widetext}
We note that the second term of Eq. (\ref{Gk_1}) is shown to vanish by use of Eqs. (\ref{G0}), (\ref{SigR0Del}), (\ref{SigA0}), and (\ref{Gra}). 
This is an expected result, where the steady-state does not depend on the initial distribution of the system $f(\omega)$ contained in $G^{0{\rm K}}$.

From the (1,2)-component of Eqs. (\ref{SigR0}), (\ref{SigR0Del}), and Eq. (\ref{Gk}), the self-consistent condition
\begin{eqnarray}
\frac{1}{U}= \sum_{\bm p}\int {d\omega \over \pi}
\frac{
F_{-}(\omega)\omega
+
F_{+}(\omega)[\xi_{\bm p}+i\gamma]
}
{[(\omega - E_{\bm p})^2 + \gamma^2]
[(\omega + E_{\bm p})^2 + \gamma^2]
},
\nonumber\\
\label{GAP}
\end{eqnarray}
is obtained.
One can also derive the number of particles $N=\sum_{\bm p,\sigma}\avg{c^\dagger_{\bm p,\sigma}c_{\bm p,\sigma}}=-2i\sum_{\bm p}[G^<(\bm p,\omega)]_{11}$ from Eqs. (\ref{Gra}) and (\ref{Gk}) as
 (where $G^{<}=(-G^{\rm R}+G^{\rm A}+G^{\rm K})/2$), 
\begin{eqnarray}
   &&
   N = 2 \sum_{\bm p} \int \frac{d\omega}{\pi} 
   \nonumber\\
   &&\times
      \frac{    \big[ (\omega + \xi_{\bm p})^2 + \gamma^2 \big] \gamma_{\rm b} f_{\rm b}(\omega) 
    + \Delta_0^2 \gamma \big[  \big ( 1 - (\gamma_{\rm b}/\gamma)f_{\rm b}(-\omega) \big ) \big]}
             { [(\omega - E_{\bm p})^2 + \gamma^2] [(\omega + E_{\bm p})^2 + \gamma ^2]}.
\nonumber\\
\label{NUM}
\end{eqnarray}
These equations (\ref{GAP}), (\ref{NUM}) can be regarded as a non-equilibrium extention of the gap and the number equations (\ref{GAPeq}) and (\ref{NUMeq}) in the BCS-Leggett theory, respectively\cite{Hanai2016}.
By solving these equations (\ref{GAP}), (\ref{NUM}) self-consistently, we determine the quantities $(\Delta_0,\mu_{\rm b},\mu)$ for a given parameter set $(a_s,N,\gamma,\kappa)$.
Note that the non-equilibrium steady-state gap equation (\ref{GAP}) is complex, making it possible to determine three quantities $(\Delta_0,\mu_{\rm b},\mu)$ from two equations (\ref{GAP}) and (\ref{NUM}).

Our formalism recovers the conventional equilibrium limit, by taking $\gamma_{\rm v}/\gamma_{\rm b}=0$ with $\gamma=\gamma_{\rm b}+\gamma_{\rm v}\rightarrow 0^+$.
This is the limit where the system is in the chemical equilibrium to the bath, since the system decouples from the vacuum.
This is seen from the imaginary part of the gap equation (\ref{GAP}), where it is satisfied when the chemical equilibrium between the bath and the system is achieved, i.e., $\mu_{\rm b}=\mu$.
In this situation, the real part of the gap equation (\ref{GAP}) and the number equation (\ref{NUM}) reduces to that of the equilibrium case (\ref{GAPeq}) and (\ref{NUMeq}), respectively. 

Once the steady-state exciton-BEC solution are determined, we can analyze the stability of the obtained steady-state, by studying the equation of motion (EOM) of a small fluctuation 
$\delta\Delta({\bm r},t)\equiv\Delta({\bm r},t)-\Delta_0$
around the order parameter $\Delta(\bm r,t)$ from the steady-state value $\Delta_0$\cite{Anderson1958,Schrieffer}. 
Note that $\delta\Delta({\bm r},t)$ can also be written as $\delta\Delta(\bm r,t)=(\Delta_0+\delta|\Delta(\bm r,t)|)e^{i\delta\theta(\bm r,t)}-\Delta_0 \simeq \delta|\Delta(\bm r,t)|+i\Delta_0\delta\theta(\bm r,t)$, 
where $\delta|\Delta(\bm r,t)|$ and $\delta\theta(\bm r,t)$ describe amplitude and phase fluctuations, respectively.
Setting $\delta|\Delta(\bm r,t)|=\delta|\Delta(\bm q,\omega)| e^{i\bm q\cdot \bm r -i\omega t} ,\delta\theta(\bm r,t)=\delta\theta(\bm q,\omega) e^{i\bm q\cdot \bm r -i\omega t} $, we obtain the linearized EOM from the Dyson's equation (\ref{Dyson}) as,
\begin{eqnarray}
M(\bm q,\omega)
\left(
\begin{array}{c}
\delta|\Delta(\bm q,\omega)|  \\
-\Delta_0\delta\theta(\bm q,\omega) 
\end{array}
\right)
=
\left(
\begin{array}{c}
0  \\
0 
\end{array}
\right),
\label{EOM}
\end{eqnarray}
where
\begin{eqnarray}
M(\bm q,\omega)
=
\left(
\begin{array}{cc}
\frac{2}{U}-\Pi_{11}(\bm q,\omega) &               -\Pi_{12}(\bm q,\omega)  \\
              -\Pi_{21}(\bm q,\omega) & \frac{2}{U}-\Pi_{22}(\bm q,\omega)
\end{array}
\right).
\label{M}
\end{eqnarray}
Here, the correlation function is given by,
\begin{eqnarray}
&&\Pi_{s,s'}(\bm q,\omega)=
\frac{i}{2}
\sum_{\bm p}\int_{-\infty}^{\infty}\frac{d\nu}{2\pi}
\nonumber\\
&&\times
{\rm Tr} 
\bigg[
\tau_s
G^{\rm R} \Big( \bm p+\frac{\bm q}{2},\nu+\frac{\omega}{2}\Big)
\tau_{s'}
G^{\rm K}\Big( \bm p - \frac{\bm q}{2},\nu-\frac{\omega}{2}\Big)
\nonumber\\
&&\ \ +
\tau_s
G^{\rm K} \Big( \bm p+\frac{\bm q}{2},\nu+\frac{\omega}{2}\Big)
\tau_{s'}
G^{\rm A}\Big( \bm p - \frac{\bm q}{2},\nu-\frac{\omega}{2} \Big)
\bigg],
\end{eqnarray}
where $\Pi_{11}(\bm q,\omega)$ and $\Pi_{22}(\bm q,\omega)$ describe amplitude and phase fluctuations of an exciton-BEC, respectively, and $\Pi_{12}(\bm q,\omega), \Pi_{21}(\bm q,\omega)$ are their couplings. 
We summarized the derivation of Eq. (\ref{EOM}) in Appendix \ref{AppEOM}.

The EOM in Eq. (\ref{EOM}) has a nontrivial solution $(\delta|\Delta(\bm q,\omega)|,-\Delta_0\delta\theta(\bm q,\omega))\neq(0,0)$, when
\begin{equation}
{\rm det} [M(\bm q,\omega=\omega_{\bm q})]=0,
\label{modeeq}
\end{equation}
which gives a (complex) mode dispersion $\omega_{\bm q}$. 
In the Bose-condensed phase, a gapless solution $\omega_{\bm q}=0$ at ${\bm q}\rightarrow 0$ of Eq. (\ref{modeeq}), or the so-called Nambu-Goldstone (NG) mode, associated with the phase fluctuations
\begin{eqnarray}
\left(
\begin{array}{c}
\delta|\Delta(\bm q\rightarrow\bm 0, \omega=0)| \\
-\Delta_0\delta\theta(\bm q\rightarrow\bm 0,\omega=0)
\end{array}
\right)
\propto
\left(
\begin{array}{c}
0 \\
1
\end{array}
\right),
\end{eqnarray}
exists (Goldstone's theorem\cite{Goldstone1961}), even in the non-equilibrium steady-state.
This can be shown from the so-called Thouless criterion\cite{Thouless1960},
\begin{eqnarray}
M_{22}(\bm q\rightarrow\bm 0,\omega=0)
&=&
{\rm Re}
\bigg[
\frac{2}{U}+\frac{i}{\Delta_0}\sum_{\bm k}\int \frac{d\omega}{2\pi}
G_{12}^{\rm K}(\bm k,\omega)
\bigg]
\nonumber\\
&=&0,
\label{ThoulessRe}
\\
M_{12}(\bm q\rightarrow\bm 0,\omega=0)
&=&
{\rm Im}
\bigg[
\frac{2}{U}+\frac{i}{\Delta_0}\sum_{\bm k}\int \frac{d\omega}{2\pi}
G_{12}^{\rm K}(\bm k,\omega)
\bigg]
\nonumber\\
&=&0.
\label{ThoulessIm}
\end{eqnarray}
In deriving Eqs. (\ref{ThoulessRe}) and (\ref{ThoulessIm}), we have used the relations (which can be derived by use of Eqs. (\ref{Gra}), (\ref{Gk}))
\begin{eqnarray}
{\rm Re} \bigg[-i\frac{G^{\rm K}_{12}(\bm k,\omega)}{\Delta_0} \bigg]
&=&
\frac{i}{2}
{\rm Tr}\Big[
\tau_2 G^{\rm R}(\bm k,\omega) \tau_2 G^{\rm K}(\bm k,\omega)
\nonumber\\
&& \ \ 
+
\tau_2 G^{\rm K}(\bm k,\omega) \tau_2 G^{\rm A}(\bm k,\omega)
\Big],
\label{G12Del}
\\
{\rm Im} \bigg[-i\frac{G^{\rm K}_{12}(\bm k,\omega)}{\Delta_0} \bigg]
&=&
\frac{i}{2}
{\rm Tr}\Big[
\tau_1 G^{\rm R}(\bm k,\omega) \tau_2 G^{\rm K}(\bm k,\omega)
\nonumber\\
&& \ \
+
\tau_1 G^{\rm K}(\bm k,\omega) \tau_2 G^{\rm A}(\bm k,\omega)
\Big],
\end{eqnarray}
in the first equality, and the gap equation (\ref{GAP}) in the second.

Noting that $\delta\Delta(\bm r,t)\propto e^{{\rm Im}[\omega_{\bm q}]t}$, one finds that an exponential growth of fluctuations occurs when ${\rm Im}[\omega_{\bm q}]>0$. 
This is the condition for dynamical instability in our approach.
In practice, we determine the dynamical instability in the following way. Since we know on physical grounds that ${\rm Im}[\omega_{\bm q}]<0$ is always satisfied at large $|\bm q|$, there exists at least one momentum ($\equiv {\tilde {\bm q}}$) satisfying ${\rm Im}[{\tilde \omega}_{\tilde {\bm q}}]=0$ if the state is dynamically unstable (${\rm Im}[\omega_{\bm q}]>0$). 
Thus, to examine the stability of an exciton-BEC, we conveniently look for a real solution of Eq. (\ref{modeeq}) with $|\tilde {\bm q}|>0$.

\section{Non-equilibrium-induced dynamical instability in the BCS-BEC crossover region}\label{sec_result}

\begin{figure}
\begin{center}
\includegraphics[width=0.65\linewidth,keepaspectratio]{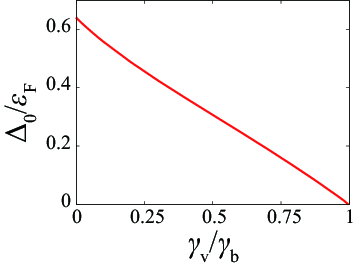}
\end{center}
\caption{(Color online) Calculated order parameter $\Delta_0$ in the unitarity limit $(k_{\rm F}a_s)^{-1}=0$, in terms of the decay rate $\gamma_{\rm v}$. We set $\gamma_{\rm b}=10^{-3}\varepsilon_{\rm F}$.}
\label{fig_del}
\end{figure}

Figure \ref{fig_del} shows the calculated order parameter $\Delta_0$ as a function of the decay rate $\gamma_{\rm v}$ at the unitarity limit $(k_{\rm F}a_s)^{-1}=0$ (which corresponds to an intermediate coupling strength).
As expected, as the decay rate $\gamma_{\rm v}$ increases, the order parameter $\Delta_0$ is suppressed by non-equilibrium effects.
At a certain decay rate $\gamma_{\rm v}/\gamma_{\rm b}\sim 1$, a transition to the normal state occurs ($\Delta_0=0$).

\begin{figure}
\begin{center}
\includegraphics[width=0.75\linewidth,keepaspectratio]{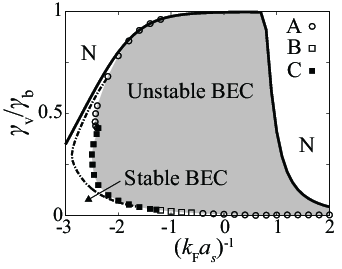}
\end{center}
\caption{
(Color online)  
Steady-state phase diagram of a driven-dissipative electron-hole gas, in terms of damping rate $\gamma_{\rm v}$ and the interaction strength $(k_{\rm F}a_s)^{-1}$. 
We set $\gamma_{\rm b}=10^{-3}\varepsilon_{\rm F}$ (where $\varepsilon_{\rm F}=k_{\rm F}^2/(2m)$ is the Fermi energy). 
The solid line is the boundary between the normal phase where $\Delta_0=0$ (``N''), and the exciton-BEC phase where $\Delta_0>0$ solution exists. 
In the exciton-BEC phase, dynamical instabilities occur at the boundaries ``A", ``B", and ``C", where the mode $\omega_{\bm q}$ starts to be associated with a negative damping rate ${\rm Im}[\omega_{\bm q}]>0$ at (A) ${\bm q}=0$, ${\rm Re}[\omega_{\bm q}]=0$, (B) $|\bm q| \ne 0$, ${\rm Re}[\omega_{\bm q}]=0$, and (C) $|\bm q| \ne 0$, ${\rm Re}[\omega_{\bm q}]\ne 0$. 
The exciton-BEC phase with $\Delta_0$ is thus unstable in the shaded region (``Unstable BEC''), while the rest of the exciton-BEC phase is stable (``Stable BEC'').
On the right side of the dashed-dotted line, a gapped mode appears\cite{NOTE}. 
}
\label{fig_phase}
\end{figure}

Our principal results are captured in Fig. \ref{fig_phase}, which shows the steady-state phase diagram of an interacting electron-hole gas.
The solid line shows the boundary between the normal phase (where $\Delta_0=0$) and the exciton-BEC phase (where $\Delta_0>0$ solution exists), and in the shaded region of the exciton-BEC phase, a dynamical instability takes place.
Our results indicate that a uniform steady-state exciton-BEC is unstable in a wide parameter region of the phase diagram, implying that non-equilibrium effects may become a large obstacle for an experimental realization of the exciton-BEC phase\cite{NOTEgammav}.

\begin{figure*}
\begin{center}
\includegraphics[width=0.75\linewidth,keepaspectratio]{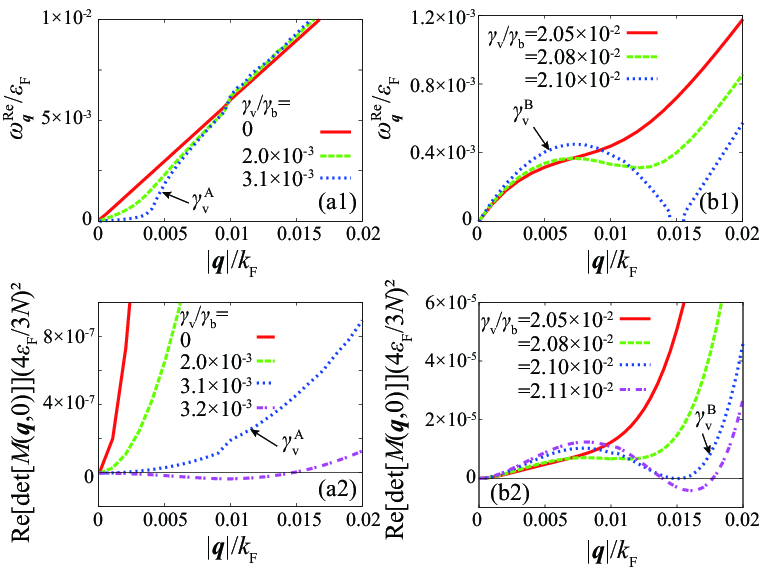}
\end{center}
\caption{(Color online) (a1), (b1) Calculated dispersion $\omega_{\bm q}^{\rm Re}$. (a2), (b2) Calculated ${\rm Re}[{\rm det}M(\bm q,\omega=0)]$. (a1), (a2) $(k_{\rm F}a_s)^{-1}=+1$. The boundary ``A'' is at $\gamma_{\rm v}^{\rm A}/\gamma_{\rm b}=3.1\times10^{-3}$ (dotted line). (b1), (b2) $(k_{\rm F}a_s)^{-1}=-1$. The boundary ``B'' is at $\gamma_{\rm v}^{\rm B}/\gamma_{\rm b}=2.10\times10^{-2}$ (dotted line). For all the figures, we set $\gamma_{\rm b}=10^{-3}\varepsilon_{\rm F}$.}
\label{fig_detM}
\end{figure*}

At present, it is unclear what happens after the dynamical instability takes place, which remains as our future work. 
However, we can still obtain some implications from the character of the instability. 
In this regard, we find that different types of dynamical instabilities take place at the boundaries ``A'', ``B'', and ``C'' in Fig. \ref{fig_phase}. 
Figure \ref{fig_detM}(a1) shows a solution of 
\begin{eqnarray}
{\rm Re}[{\rm det}[M(\bm q,\omega_{\bm q}^{\rm Re})]]=0,
\label{dispersionRe}
\end{eqnarray}
in the vicinity of the boundary ``A'', which gives the dispersion $\omega_{\bm q}^{\rm Re}$ of a collective mode with its linewidth neglected.
Here, as discussed in Sec. \ref{sec_formalism}, a gapless NG mode is obtained as a result of the Thouless criterion (\ref{ThoulessRe}), (\ref{ThoulessIm}).
We clearly see that the sound velocity of the NG mode gradually decreases as the decay rate $\gamma_{\rm v}$ increases, until it reaches zero at the boundary ``A'' (dotted line), leading to a dynamical instability.
We plotted in Fig. \ref{fig_detM}(a2) the $|\bm q|$ dependence of ${\rm Re}[{\rm det}M(\bm q,\omega)]$ at $\omega=0$.
Noting that ${\rm Im}[{\rm det}M(\bm q,\omega)]=0$ is always satisfied at $\omega=0$, ${\rm Re}[{\rm det}M(\tilde{\bm q},\omega=0)]=0$ means that a pole exists at $\tilde{\bm q}$.
Starting from the stable region (solid and dashed line), as $\gamma_{\rm v}$ increases,
the curvature of ${\rm Re}[{\rm det}M(\bm q,\omega=0)]$ at small $|\bm q|$ gradually decreases,  which approaches zero at the boundary ``A'' (dotted line). 
Beyond this boundary (dashed-dotted line), ${\rm Re}[{\rm det}M(\bm q,\omega=0)]$ becomes downward-convex at small $|\bm q|$, which leads to the appearance of a pole starting from $\tilde{\bm q}\rightarrow 0$.
This shows that, at the boundary ``A'', the instability condition ${\rm Im}[\omega_{\bm q}]>0$ is obtained in the long wavelength and low energy limit (${\rm Re}[\omega_{\bm q}]=0$), implying a collapse of exciton-BEC.

\begin{figure}
\begin{center}
\includegraphics[width=0.8\linewidth,keepaspectratio]{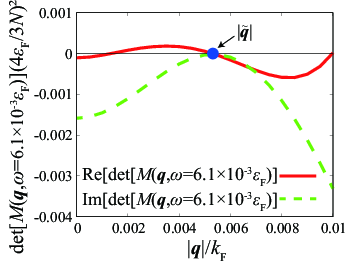}
\end{center}
\caption{(Color online) Calculated ${\rm det}M(\bm q,\omega=6.1\times 10^{-3}\varepsilon_{\rm F})$ at the boundary ``C'' ($\gamma_{\rm v}^{\rm C}/\gamma_{\rm b}=7.4\times 10^{-2}$). The interaction strength is $(k_{\rm F}a_s)^{-1}=-2$, and we set $\gamma_{\rm b}=10^{-3}\varepsilon_{\rm F}$.}
\label{fig_detM_C}
\end{figure}

At the boundaries ``B'' and ``C'', on the other hand, instability occurs at finite momentum $|\bm q|$.
The dispersion $\omega_{\bm q}^{\rm Re}$ determined from Eq. (\ref{dispersionRe}) in the vicinity of the boundary ``B'' is plotted in Fig. \ref{fig_detM}(b1), showing a dispersion with a local minimum (dashed line), similar to a roton minimum in a superfluid ${}^4$He. 
The mode touches $\omega=0$ at the boundary ``B'' (dotted line), which leads to a dynamical instability at $|\bm q|\ne 0$.
There, as shown in Fig. \ref{fig_detM}(b2), the condition ${\rm Im}[\omega_{\bm q}]>0$ is realized at $|\bm q|\ne 0,{\rm Re}[\omega_{\bm q}]=0$ (dotted line).
We also find regimes (boundary ``C'') where dynamical instabiliity occurs at $|\bm q|\ne 0,{\rm Re}[\omega_{\bm q}]\ne 0$, as shown in Fig. \ref{fig_detM_C}.
Type ``B'' implies instability being accompanied by a pattern formation, and type ``C'' implies a transition to a non-steady-state with space modulation.

To grasp the essence of the above-mentioned dynamical instabilities, we now derive the (approximate) hydrodynamic equations (\ref{tGL}) and (\ref{continuity}) by expandng $M(\bm q,\omega)$ in Eq. (\ref{EOM}) in terms of $\bm q$ and $\omega$. 
The form of this expansion can be restricted from symmetry considerations;
firstly, from the inversion symmetry, $M(\bm q,\omega)=M(-\bm q,\omega)$.
Secondly, since $\delta|\Delta(\bm r,t)|$ and $\delta\theta(\bm r,t)$ are real, 
\begin{eqnarray}
M(\bm q,\omega)=M^*(\bm q,-\omega).
\label{Mreal}
\end{eqnarray}
Finally, the Goldstone's theorem (Eqs. (\ref{ThoulessRe}) and (\ref{ThoulessIm})) assures $M_{12}(\bm q\rightarrow\bm 0,\omega=0)=M_{22}(\bm q\rightarrow \bm 0,\omega=0)=0$.
These yield, up to $O(\bm q^2,\omega)$, 
\begin{eqnarray}
M(\bm q,\omega)
=
\left(
\begin{array}{cc}
\alpha_{11}+\beta_{11}\bm q^2-i\eta_{11}\omega & -\beta_{12}\bm q^2 + i\eta_{12}\omega  \\
\alpha_{21}+\beta_{21}\bm q^2 -i\eta_{21}\omega & \beta_{22}\bm q^2 - i\eta_{22}\omega  
\end{array}
\right),
\nonumber\\
\label{Mslow}
\end{eqnarray}
where the coefficients $\alpha_{ij},\beta_{ij},$ and $\eta_{ij}$ ($i,j=1,2$) are real numbers.
As a result, the form of the determinant can be determined (up to $O(\bm q^2,\omega)$) as, 
\begin{eqnarray}
{\rm det}M(\bm q,\omega)=A(\omega^2+i\Gamma\omega-c\bm q^2),
\end{eqnarray}
giving rise to the so-called diffusive Goldstone's mode\cite{Wouters2007,Szymanska2006},
\begin{eqnarray}
\omega_{\bm q}=-i\frac{\Gamma}{2}\pm\sqrt{c\bm q^2-\frac{\Gamma^2}{4}}.
\end{eqnarray}
The real numbers $A,c$ and $\Gamma$ are determined from coefficients $\alpha_{ij},\beta_{ij}$, and $\eta_{ij}$ in Eq. (\ref{Mslow}).
We note that the obtained EOM breaks the time reversal symmetry ($\omega\rightarrow -\omega,|\delta\Delta(\bm q,\omega)|\rightarrow|\delta\Delta(\bm q,-\omega)|,\delta\theta(\bm q,\omega)\rightarrow -\delta\theta(\bm q,-\omega)$), due to the tunneling from the bath and the vacuum through $\gamma_{\rm b}$ and $\gamma_{\rm v}$. 

However, in the limit where $\gamma_{\rm b}$ and $\gamma_{\rm v}$ are both small (i.e., $\gamma=\gamma_{\rm b}+\gamma_{\rm v}\rightarrow 0^+$ with $\gamma_{\rm v}/\gamma_{\rm b}$ fixed), it can be shown that Eq. (\ref{EOM}) further reduces to 
\begin{eqnarray}
(\alpha_{11}-\beta_{11}\nabla^2)\delta|\Delta(\bm r,t)|^2 
+ 2\Delta_0^2\eta_{12}\partial_t \delta\theta(\bm r,t)
&=&0,  
\label{tGL_1}\\
\eta_{12}\partial_t\delta|\Delta(\bm r,t)|^2+2\Delta_0^2\beta_{22}\nabla^2\delta\theta(\bm r,t) 
&=&0,
\label{continuity_1}
\end{eqnarray}
where the time reveral symmetry is recovered.
Here, we have used Eq. (\ref{Mreal}), as well as the relations that holds in this limit,
\begin{eqnarray}
\Pi_{11}(\bm q,\omega)& =& \Pi_{11}(\bm q,-\omega),\\
\Pi_{12}(\bm q,\omega)&=&-\Pi_{21}(\bm q,\omega),\\
\Pi_{22}(\bm q,\omega)& =& \Pi_{22}(\bm q,-\omega),\\
\Pi_{12}(\bm q,\omega=0)&=&\Pi_{21}(\bm q,\omega=0)=0,
\end{eqnarray}
which can be obtained from relations (Take the $\gamma\rightarrow 0^+$ limit of Eqs. (\ref{Gra}), (\ref{Gk}).),
\begin{eqnarray}
G^{\rm R}_{11}(\bm k,-\omega)
&=&[G^{\rm A}_{11}(\bm k,-\omega)]^*
\nonumber\\
&=&-G^{\rm A}_{22}(\bm k,\omega)
=-[G^{\rm R}_{22}(\bm k,\omega)]^*,
\label{gr11momega}
\\
G^{\rm R}_{22}(\bm k,-\omega)
&=&[G^{\rm A}_{22}(\bm k,-\omega)]^*
\nonumber\\
&=&-G^{\rm A}_{11}(\bm k,\omega)
=-[G^{\rm R}_{11}(\bm k,\omega)]^*,
\label{gk11momega}
\\
G^{\rm R}_{12}(\bm k,-\omega)&=&G^{\rm A}_{12}(\bm k,\omega),
\label{gr12momega}
\\
G^{\rm K}_{11}(\bm k,-\omega)
&=&-[G^{\rm K}_{11}(\bm k,-\omega)]^*
\nonumber\\
&=&-G^{\rm K}_{22}(\bm k,\omega)
=[G^{\rm K}_{22}(\bm k,\omega)]^*,
\\
G^{\rm K}_{12}(\bm k,-\omega)&=&G^{\rm K}_{12}(\bm k,\omega),
\label{gk12momega}
\end{eqnarray}
and
\begin{eqnarray}
G^{\rm R/A/K}_{12}(\bm k,\omega)=G^{\rm R/A/K}_{21}(\bm k,\omega).
\label{Grak12}
\end{eqnarray}
Note that this limit does $\textit{not}$ necessarily correspond to the equilibrium limit, since the ratio between the thermalization rate $\gamma_{\rm b}$ and the decay rate $\gamma_{\rm v}$ is kept finite (unless $\gamma_{\rm v}/\gamma_{\rm b}=0$).
Using the fact that the Bose-condensate fraction $n_0=\sum_{\bm p}|\avg{c_{-\bm p,{\rm h}}c_{\bm p,{\rm e}}}|^2$ is proportional to $\Delta_0^2$, we obtain the hydrodynamic equations (\ref{tGL}) and (\ref{continuity}) of the fluctuations of the condensate fraction $\delta n_0(\bm r,t)$ around the steady-state value $n_0$ and the phase fluctuations $\delta\theta(\bm r,t)$.
As already pointed out in the introduction, by comparing the obtained hydrodynamic equations (\ref{tGL}), (\ref{continuity}) to the Gross-Pitaevskii (GP) equations (\ref{tGLGP}), (\ref{continuityGP}), we find that $\alpha_{11}/\eta_{12}$ may be regarded as an effective interaction $U_{\rm eff}$ between excitons multiplied by the condensate fraction $n_0$ of excitons.

\begin{figure*}
\begin{center}
\includegraphics[width=0.75\linewidth,keepaspectratio]{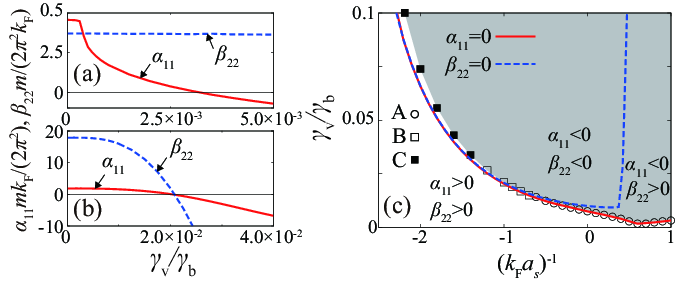}
\end{center}
\caption{
(Color online)  
(a), (b) Coefficients $\alpha_{11}$ and $\beta_{22}$ (a) in the BEC regime $(k_{\rm F}a_s)^{-1}=+1$ where type ``A'' instability takes place, and (b) in the BCS regime $(k_{\rm F}a_s)^{-1}=-1$ where type ``B'' instability takes place.
(c) The solutions of $\alpha_{11}=0$ and $\beta_{22}=0$ in the steady-state phase diagram. 
As in Fig. \ref{fig_phase}, dynamical instabilities occur at the boundaries ``A", ``B", and ``C". 
The shaded area represent the dynamical unstable region, while the white area represents the region where the uniform steady-state BEC is stable.
We briefly note that dynamical instabilities ``B'' and ``C'' in the BCS regime occur slightly above the boundaries $\alpha_{11}=0$ and $\beta_{22}=0$, while type ``A'' occurs (almost) at $\alpha_{11}=0$.
}
\label{fig_phase_inset}
\end{figure*}

The coupled equations (\ref{tGL_1}), (\ref{continuity_1}) gives the determinant
\begin{eqnarray}
{\rm det}M(\bm q,\omega)
=\beta_{22}\bm q^2(\alpha_{11}+\beta_{11}\bm q^2) 
- \eta_{12}^2\omega^2,
\label{detM_gam0}
\end{eqnarray}
and the mode dispersion as,
\begin{eqnarray}
\omega_{\bm q}=\pm\frac{\sqrt{\beta_{22} \bm q^2(\alpha_{11}+\beta_{11}\bm q^2)}}{\eta_{12}}.
\label{mode}
\end{eqnarray}
When all the coefficients are positive, Eq. (\ref{mode}) gives a stable sound mode.
However, one sees in Fig. \ref{fig_phase_inset}(a) that $\alpha_{11}$ becomes negative in the BEC regime when $\gamma_{\rm v}$ exceeds a certain value.
According to Eq. (\ref{mode}), this leads to a dynamical instability.
(Note however that Eq. (\ref{mode}) is only valid at $\gamma=\gamma_{\rm b}+\gamma_{\rm v}\rightarrow 0^+$.)
As seen in Fig. \ref{fig_phase_inset}(c), since this sign change occurs (almost) at the boundary ``A'' in the BEC side ($(k_{\rm F}a_s)^{-1}\gesim 0$), this instability is considered to be caused by an effective \textit{attractive} interaction between excitons.
We briefly note that this situation corresponds to the case illustrated in Fig. \ref{fig_instability}(b).

\begin{figure*}
\begin{center}
\includegraphics[width=0.55\linewidth,keepaspectratio]{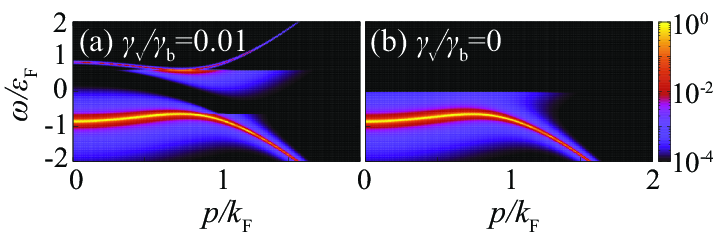}
\end{center}
\caption{(Color online) Calculated intensity of the occupied spectral weight $L(\bm p,\omega)$ in the unitarity limit $(k_{\rm F}a_s)^{-1}=0$. The spectral intensity is normalized by $\varepsilon_{\rm F}^{-1}$. (a) $\gamma_{\rm v}=0.01\gamma_{\rm b}$ (non-equilibrium case). (b) $\gamma_{\rm v}=0$ (equilibrium case). We set $\gamma_{\rm b}=10^{-3}\varepsilon_{\rm F}$.}
\label{fig_ospc}
\end{figure*}

This anomaly ``A'' is attributed to pair-dissociation induced by pumping and decay.
Figure \ref{fig_ospc} shows the occupied spectral weight function 
$L(\bm p,\omega)=\int_{-\infty}^{\infty}dt e^{i\omega t}\avg{c^\dagger_{\bm p}(t)c_{\bm p}(0)}$, 
which describes the occupancy of single-particle density of states.
We clearly see from Fig. \ref{fig_ospc}(a) that the pumping supplies carriers to the upper branch of the single-particle excitation spectrum ($\omega=E_{\bm p}$, where $E_{\bm p}=\sqrt{(\varepsilon_{\bm p}-\mu)^2+\Delta_0^2}$), as opposed to the equilibrium case (Fig. \ref{fig_ospc}(b)) where only the lower branch ($\omega=-E_{\bm p}$) is occupied.
This indicates that pair-breaking occurs in a non-equilibrium situation\cite{Hanai2016,Yamaguchi2012}.

Keeping this in mind, the above-mentioned attractive interaction $U_{\rm eff}$ can be shown to be caused by this non-equilibrium induced pair-breaking effects, by expanding ($\gamma\rightarrow 0^+$)
\begin{eqnarray}
&&
\alpha_{11}
\xrightarrow{\gamma\rightarrow 0^+}
\frac{2}{U}
-2i\sum_{\bm k}\int \frac{d\nu}{2\pi}
\nonumber\\
&&\times
\Bigg[
\bigg[
G_{12}\Big( \bm k+\frac{\bm q}{2},\nu+\frac{\omega}{2} \Big)
G_{12}\Big( \bm k-\frac{\bm q}{2},\nu-\frac{\omega}{2}\Big)
\nonumber\\
&&
+
G_{11}\Big( \bm k+\frac{\bm q}{2},\nu+\frac{\omega}{2} \Big)
G_{22}\Big( \bm k-\frac{\bm q}{2},\nu-\frac{\omega}{2}\Big)
\bigg]^{\rm K}
\Bigg]_{\bm q\rightarrow 0,\omega=0}
\nonumber\\
&&=
-4i\sum_{\bm k}\int \frac{d\nu}{2\pi}
\nonumber\\
&&\times
\Bigg[
\bigg[
G_{12}\Big( \bm k+\frac{\bm q}{2},\nu+\frac{\omega}{2} \Big)
G_{12}\Big( \bm k-\frac{\bm q}{2},\nu-\frac{\omega}{2}\Big)
\bigg]^{\rm K}
\Bigg]_{\bm q\rightarrow 0,\omega=0},
\nonumber\\
\label{alpha11_1}
\end{eqnarray}
in terms of $\Delta_0$, where Eq. (\ref{Grak12}) is used in obtaining the first equality, and the Thouless criterion (\ref{ThoulessRe}) and Eq. (\ref{G12Del}) in the second. 
Here, we have introduced a short-hand notation,
\begin{eqnarray}
&&[\hat G \hat G \hat G\cdots \hat G \hat G ]^{\rm R}=G^{\rm R}G^{\rm R}G^{\rm R}\cdots G^{\rm R}G^{\rm R}
\label{shortR}
\\ 
&&[\hat G \hat G \hat G\cdots \hat G \hat G ]^{\rm A}=G^{\rm A}G^{\rm A}G^{\rm A}\cdots G^{\rm A}G^{\rm A}
\label{shortA}
\\ 
&&[\hat G \hat G \hat G\cdots \hat G \hat G ]^{\rm K}
=
G^{\rm R}G^{\rm R} \cdots G^{\rm R}G^{\rm K}
+G^{\rm R}G^{\rm R} \cdots G^{\rm R} G^{\rm K}G^{\rm A}\nonumber\\
&&\ \ \ +\cdots
+G^{\rm R}G^{\rm R} \cdots G^{\rm R} G^{\rm K}G^{\rm A}G^{\rm A}\cdots G^{\rm A}
+\cdots
\nonumber\\
&& \ \ \ \ +
G^{\rm R}G^{\rm K}G^{\rm A}G^{\rm A}\cdots G^{\rm A}
+G^{\rm K}G^{\rm A}G^{\rm A}\cdots G^{\rm A}.
\label{shortK}
\end{eqnarray}
This expansion is justified in the BEC regime, where $\Delta_0$ can be regarded to be small compared to the binding energy (i.e., $\Delta_0\ll |\mu|\simeq E_{\rm b}/2$)\cite{NOTEDeltaexpand}. 
\begin{widetext}
In this regime, the single-particle Green's function $\hat G$ can be expanded as,
\begin{eqnarray}
G^{\rm R}(\bm k,\omega )
&=&
[(\omega + i\gamma) \bm 1 - \xi_{\bm k}\tau_3+\Delta_0 \tau_1]^{-1}
=
\tilde G^{0{\rm R}}(\bm k,\omega)
-[\hat{\tilde G}^0(\bm k,\omega)\tau_1 \hat{\tilde G}^0(\bm k,\omega)]^{\rm R}\Delta_0
\nonumber\\
&+& 
[\hat{\tilde G}^0(\bm k,\omega)\tau_1 \hat{\tilde G}^0(\bm k,\omega)\tau_1 \hat{\tilde G}^0(\bm k,\omega)]^{\rm R} \Delta_0^2
 +\cdots,
 \label{GrDelseries}
 \nonumber\\
 \\
G^{\rm A}(\bm k,\omega)&=& [G^{\rm R}(\bm k,\omega)]^\dagger,
\label{GaDelseries}
\end{eqnarray}
and
\begin{eqnarray}
G^{\rm K}(\bm k,\omega)
&=&
G^{\rm R}(\bm k,\omega)\Sigma^{\rm K}_0(\omega) G^{\rm A}(\bm k,\omega)
=
\tilde G^{0{\rm K}}(\bm k,\omega)
-[\hat{\tilde G}^0(\bm k,\omega)\tau_1 \hat{\tilde G}^0(\bm k,\omega)]^{\rm K}
\Delta_0
\nonumber\\
&+&
[\hat{\tilde G}^0(\bm k,\omega)\tau_1 \hat{\tilde G}^0(\bm k,\omega)\tau_1 \hat{\tilde G}^0(\bm k,\omega)]^{\rm K}
\Delta_0^2
+\cdots,
\nonumber\\
\label{GkDelseries}
\end{eqnarray}
where 
\begin{eqnarray}
\hat{\tilde G}^0(\bm k,\omega)
&=&
[\hat G^0 (\bm k,\omega)- \hat \Sigma_{\rm env}(\omega)]^{-1}
\nonumber\\
&=&
\left(
\begin{array}{cc}
[\omega+i\gamma -\xi_{\bm k}\tau_3]^{-1} & - i \tau_3 (1-2f_\lambda(\omega\tau_3)) \frac{\gamma}{(\omega-\xi_{\bm k}\tau_3)^2+\gamma^2} \\
0                  & [\omega-i\gamma -\xi_{\bm k}\tau_3]^{-1}
\end{array}
\right)
\nonumber\\
&\xrightarrow{\gamma\rightarrow 0^+}&
\left(
\begin{array}{cc}
[\omega+i\delta -\xi_{\bm k}\tau_3]^{-1} & -\pi i \tau_3 (1-2f_\lambda(\omega\tau_3)) \delta(\omega-\xi_{\bm k}\tau_3) \\
0                  & [\omega-i\delta -\xi_{\bm k}\tau_3]^{-1}
\end{array}
\right).
\label{Gtildeconcrete}
\end{eqnarray}
is a Green's function with $\Delta_0=0$. 
(Note however that its distribution is modified by the bath and the vacuum distribution $f_\lambda(\omega)$). 
Substituting Eqs. (\ref{GrDelseries}), (\ref{GaDelseries}), and (\ref{GkDelseries}) into Eq. (\ref{alpha11_1}), we get
\begin{eqnarray}
\alpha_{11}
&\simeq & 
-4i\Delta_0^2 \sum_{\bm k}\int_{-\infty}^{\infty}\frac{d\nu}{2\pi}
\bigg[
\Big[
\tilde G^0_{11}(\bm k+\frac{\bm q}{2},\nu+\frac{\omega}{2})
\tilde G^0_{22}(\bm k+\frac{\bm q}{2},\nu+\frac{\omega}{2})
\tilde G^0_{11}(\bm k-\frac{\bm q}{2},\nu-\frac{\omega}{2})
\tilde G^0_{22}(\bm k-\frac{\bm q}{2},\nu-\frac{\omega}{2})
\Big]^{\rm K}
\bigg]_{\bm q\rightarrow 0,\omega=0}
\nonumber\\
&=&
-4i\Delta_0^2 \sum_{\bm k}\int_{-\infty}^{\infty}\frac{d\nu}{2\pi}
\nonumber\\
&&\times
\bigg[
\Big[
 \tilde G^0_{11}(\bm k+\frac{\bm q}{2},\nu+\frac{\omega}{2})
[\tilde G^0_{11}(-\bm k-\frac{\bm q}{2},-\nu-\frac{\omega}{2})]^* 
 \tilde G^0_{11}(\bm k-\frac{\bm q}{2},\nu-\frac{\omega}{2})
[\tilde G^0_{11}(-\bm k+\frac{\bm q}{2},-\nu+\frac{\omega}{2})]^*
\Big]^{\rm K}
\bigg]_{\bm q\rightarrow 0,\omega=0},
\label{inter_exciton_interaction}
\end{eqnarray}
where we have used Eqs. (\ref{gr11momega}) and (\ref{gk11momega}) in the last equality.
\end{widetext}

\begin{figure}
\begin{center}
\includegraphics[width=0.6\linewidth,keepaspectratio]{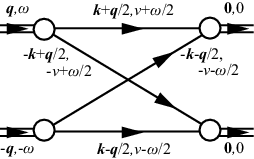}
\end{center}
\caption{(Color online) Diagramatical expression of effective interaction between excitons $U_{\rm eff}$ from an exchange process of electrons and holes. The solid line represents $\tilde G^0_{11}$, and the double solid line represents the incoming or outgoing exciton.}
\label{figinter_exciton_interaction}
\end{figure}

Equation (\ref{inter_exciton_interaction}) can be depicted diagramatically as in Fig. \ref{figinter_exciton_interaction}.
This diagram shows that the effective exciton-exciton interaction $U_{\rm eff}$ can be expressed as an exchange process of particles between the two incoming excitons\cite{Pieri2005,OhashiJPSJ}. 
By using Eq. (\ref{Gtildeconcrete}) and performing the $\nu$-integral, we obtain
\begin{eqnarray}
&&
\alpha_{11}
\simeq
\Delta_0^2
\sum_{\bm k}
\bigg[
\frac{1}{\xi_{+}\xi_{-}}
\frac{\xi_{+}+\xi_{-}}{(\xi_{+}+\xi_{-})^2-(\omega+i\delta)^2}
\nonumber\\
&&\ \ \ \ \ \ \ \ \ \ \ \ \ \ \ \ \ \ 
\times
2\Big(
1
-\frac{\gamma_{\rm b}}{\gamma}
f_{\rm b}(\xi_{+})
-\frac{\gamma_{\rm b}}{\gamma}
f_{\rm b}(\xi_{-})
\Big)
\nonumber\\
&&\ \ \ \ \ \ \ \ \ \ \ \ \ \ \ \ \ \ 
-\frac{1}{\xi_{+}\xi_{-}}
\frac{\xi_{+}-\xi_{-}}{(\xi_{+}-\xi_{-})^2-(\omega+i\delta)^2}
\nonumber\\
&&\ \ \ \ \ \ \ \ \ \ \ \ \ \ \ \ \ \ 
\times
\Big(
-\frac{2\gamma_{\rm b}}{\gamma}
\Big)
(f_{\rm b}(\xi_{+})-f_{\rm b}(\xi_{-}))
\bigg]_{\bm q\rightarrow 0,\omega=0}
\nonumber\\
&=&
\Delta_0^2
\sum_{\bm k}
\bigg[
\frac{1}{\xi_{\bm k}^3}
\Big(
1-\frac{2\gamma_{\rm b}}{\gamma}f_{\rm b}(\xi_{\bm k})
\Big)
-\frac{1}{\xi_{\bm k}^2}\frac{2\gamma_{\rm b}}{\gamma} 
\Big(
-\frac{df_{\rm b}(\xi_{\bm k})}{d\xi_{\bm k}}
\Big)
\bigg],
\nonumber\\
\label{alpha11Ueff}
\end{eqnarray}
where $\xi_\pm=\xi_{\pm\bm k+\bm q/2}$.

In the equilibrium case where dissociated pairs are absent, i.e., $f_{\rm b}(\xi_{\bm k})=0$ (Note that $\xi_{\bm k}=k^2/(2m)+|\mu|>0$.), only the first term in Eq. (\ref{alpha11Ueff}) exists. 
In this case, $\alpha_{11}$ is always positive, indicating that the effective exciton-exciton interaction $U_{\rm eff}$ is \textit{repulsive}\cite{Pieri2005,OhashiJPSJ}.
On the other hand, in the non-equilibrium case where quasiparticle excitations due to pair-dissociation effects are present ($f_{\rm b}(\xi_{\bm k})>0$), the second term in Eq. (\ref{alpha11Ueff}) arises.
Noting $-(df_{\rm b}(\xi_{\bm k})/d\xi_{\bm k}) > 0$, this term gives a negative contribution, clearly indicating that the non-equilibrium induced pair-breaking effect gives rise to an \textit{attractive} interaction. 
Our numerical analysis shows that the attractive channel from the second term can become superior to the repulsive channel from the first term, giving rise to the dynamical instability ``A''.

\begin{figure}
\begin{center}
\includegraphics[width=1\linewidth,keepaspectratio]{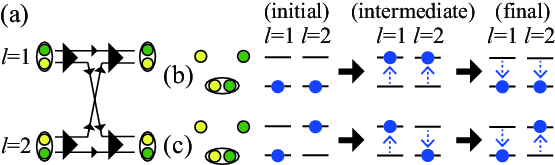}
\end{center}
\caption{(Color online) (a) Scattering process giving the effective interaction $U_{\rm eff}$. (b) Scattering process where two excitons in the initial state virtually dissociate into electrons and holes in the intermediate state. (c) Scattering process where one of the two excitons are initially dissociated. In the intermediate state, while the depairing of an exciton occurs, the electron and hole in the initial state form an exciton. In panels (b) and (c), the lower (upper) level schematically represents the exciton state (dissociated electron-hole mixture).}
\label{fig_interaction_sch}
\end{figure}

The above can be interpreted as follows.
As mentioned earlier, the exciton-exciton interaction $U_{\rm eff}$ is obtained from scattering processes shown in Fig. \ref{fig_interaction_sch}(a)\cite{Pieri2005,OhashiJPSJ}, which is accompanied by exchange of particles between two excitons. 
When two excitons labeled as $l=1$ and $l=2$ with the energies $\omega_{\rm i}^{l}$ are virtually dissociated into electrons and holes in the intermediate state (Fig. \ref{fig_interaction_sch}(b)), the resulting effective interaction involves $[\omega_{\rm i}^{l=1}-\omega_{\rm m}^{l=1}]^{-1}\times[\omega_{\rm i}^{l=2}-\omega_{\rm m}^{l=2}]^{-1}>0$, where $\omega_{\rm m}^{l}$ $(>\omega_{\rm i}^{l})$ is the energy of a dissociated exciton. The positivity of this factor means that this process gives a {\it repulsive} interaction\cite{Pieri2005,OhashiJPSJ}. On the other hand, when some excitons are dissociated in the non-equilibrium steady-state, we obtain an additional contribution to $U_{\rm eff}$ schematically shown in Fig. \ref{fig_interaction_sch}(c). In this case, starting from the the initial state with one exciton and an electron and hole, in the intermediate states, while the exciton is dissociated, the electron and hole in the initial state form an exciton. 
This process gives an \textit{attractive} contribution to $U_{\rm eff}$ being proportional to $[\omega_{\rm i}^{l=1}-\omega_{\rm m}^{l=1}]^{-1}\times[\omega_{\rm i}^{l=2}-\omega_{\rm m}^{l=2}]^{-1}<0$. 
When the contribution from the latter exceeds the former to give an attractive exciton-exciton interaction, the system becomes unstable.

\begin{figure}
\begin{center}
\includegraphics[width=0.4\linewidth,keepaspectratio]{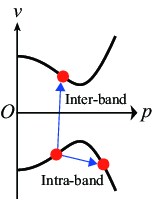}
\end{center}
\caption{(Color online) Schematic explanations for inter- and intra-band excitations. }
\label{fig_TE_QS}
\end{figure}

\begin{figure}
\begin{center}
\includegraphics[width=1\linewidth,keepaspectratio]{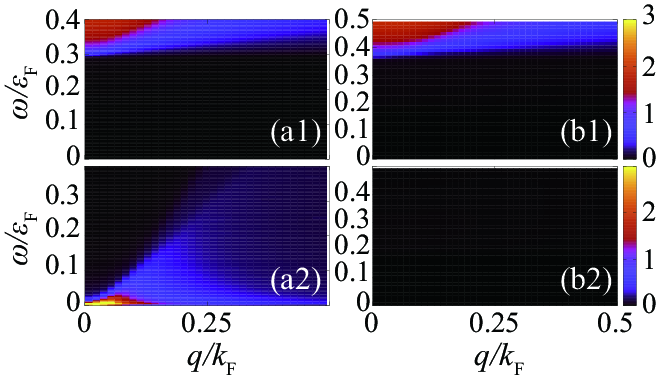}
\end{center}
\caption{(Color online) Calculated spectrum of pair-correlation function ${\rm Im}\Pi={\rm Im}\Pi^{\rm inter}+{\rm Im}\Pi^{\rm intra}$. (a1), (b1) Inter-band ${\rm Im}[\Pi^{\rm inter}_{11}(\bm q,\omega)]$ contribution. (a2), (b2) Intra-band contribution ${\rm Im}[\Pi^{\rm intra}_{11}(\bm q,\omega)]$. (a1), (a2) Non-equilibrium case $\gamma_{\rm v}/\gamma_{\rm b}=0.1$. (b1), (b2) Equilibrium case $\gamma_{\rm v}/\gamma_{\rm b}=0$. Interaction strength is set to $(k_{\rm F}a_s)^{-1}=-1$.}
\label{fig_spcPI}
\end{figure}

So far, we have shown that the pair-dissociation seen in Fig. \ref{fig_ospc}(a) plays a crucial role for the occurance of the dynamical instability ``A'' in the BEC side. 
Below, we show that in fact, the non-equilibrium induced pair-dissociation is responsible for dynamical instabilities in \textit{all} regimes, including the BCS regime where types ``B'' and ``C'' instabilities occur. 
For this purpose, it is convenient to divide the pair-correlation function $\Pi=\Pi^{\rm inter}+\Pi^{\rm intra}$ into the inter-band
\begin{eqnarray}
&&
\Pi^{\rm inter}_{s,s'}(\bm q,\omega)
=
\frac{i}{2}
\sum_{\bm p}\int_{-\infty}^{\infty}\frac{d\nu}{2\pi}
\nonumber\\
&&\times
{\rm Tr} 
\bigg[
\tau_s
G_{\rm l} \Big( \bm p+\frac{\bm q}{2},\nu+\frac{\omega}{2}\Big)
\tau_{s'}
G_{\rm u}\Big( \bm p - \frac{\bm q}{2},\nu-\frac{\omega}{2}\Big)
\nonumber\\
&&
+
\tau_s
G_{\rm u} \Big( \bm p+\frac{\bm q}{2},\nu+\frac{\omega}{2}\Big)
\tau_{s'}
G_{\rm l}\Big( \bm p - \frac{\bm q}{2},\nu-\frac{\omega}{2} \Big)
\bigg]^{\rm K},
\end{eqnarray}
and the intra-band  
\begin{eqnarray}
&&
\Pi^{\rm intra}_{s,s'}(\bm q,\omega)
=
\frac{i}{2}
\sum_{\bm p}\int_{-\infty}^{\infty}\frac{d\nu}{2\pi}
\nonumber\\
&&\times
{\rm Tr} 
\bigg[
\tau_s
G_{\rm l} \Big( \bm p+\frac{\bm q}{2},\nu+\frac{\omega}{2}\Big)
\tau_{s'}
G_{\rm l}\Big( \bm p - \frac{\bm q}{2},\nu-\frac{\omega}{2}\Big)
\nonumber\\
&&
+
\tau_s
G_{\rm u} \Big( \bm p+\frac{\bm q}{2},\nu+\frac{\omega}{2}\Big)
\tau_{s'}
G_{\rm u}\Big( \bm p - \frac{\bm q}{2},\nu-\frac{\omega}{2} \Big)
\bigg]^{\rm K},
\end{eqnarray}
contributions\cite{Wong1988,Ohashi1997} (See Fig. \ref{fig_TE_QS}.), by splitting the single-particle Green's function $\hat G=\hat G_{\rm l}+\hat G_{\rm u}$ into the lower branch ($\omega=-E_{\bm p}$) contribution $\hat G_{\rm l}$ and the upper branch ($\omega=E_{\bm p}$) contribution $\hat G_{\rm u}$ to the single-particle excitations.
We gave the concrete definition of $\hat G_{\rm l}$ and $\hat G_{\rm u}$ in Appendix \ref{App_Gul}.
Here, the inter-band contribution $\Pi^{\rm inter}$ describes effects of two-particle excitations such as molecular dissociation into two quasi-particles, which is present even in the equilibrium state.
Figures \ref{fig_spcPI}(a1) and (b1) show the spectrum ${\rm Im}\Pi^{\rm inter}_{11}(\bm q,\omega)$ in the non-equilibrium and the equilibrium cases, respectively. 
Here, we see no qualitative differences between the two cases, where both spectra exhibits an energy gap of $\omega\simeq 2\Delta_0$, i.e., the least amount of energy required to dissociate pairs.

On the other hand, the intra-band contribution $\Pi^{\rm intra}$ describes the scattering effects between quasi-particles.
In contrast to the inter-band excitations, this process only appears when quasi-particles already exist (i.e., dissociated pairs are present) in the steady-state, which is characteristic of the non-equilibrium case (See Fig. \ref{fig_ospc}.).
As a result, as seen in ${\rm Im}\Pi^{\rm intra}_{11}(\bm q,\omega)$ in Figs. \ref{fig_spcPI}(a2) and (b2) for the non-equilibrium and the equilibrium case, respectively, this quantity gives finite contribution only in the non-equilibrium case.
Note that at small $|\bm q|$, ${\rm Im}\Pi^{\rm intra}(\bm q,\omega)$ gives a remarkable contribution only at low energy $\omega\ll \Delta_0$.
This feature is quite similar to the continuum spectrum in the density fluctuations of an interacting electron gas\cite{Fetter}, which is reasonable because quasi-particle scattering can be regarded somewhat as a single-particle excitation process of ``quasi-particle density'' fluctuations.

\begin{figure*}
\begin{center}
\includegraphics[width=0.55\linewidth,keepaspectratio]{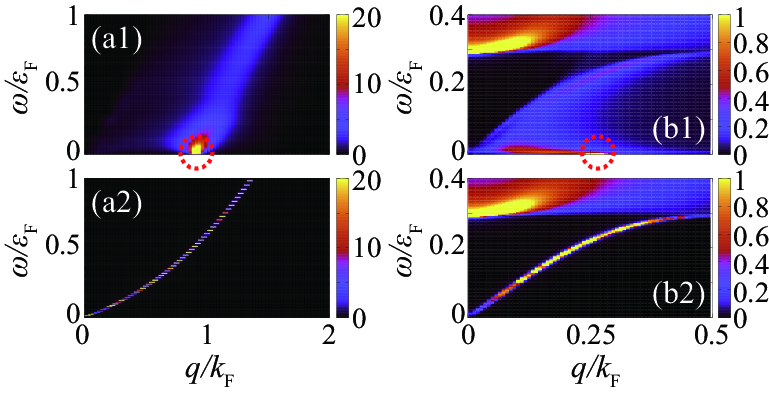}
\end{center}
\caption{(Color online) (a1), (b1) Calculated spectral function ${\rm Im}[M^{-1}(\bm q,\omega)]_{11}$.
(a2), (b2) The same as (a1) and (b1), with intra-band contribution $\Pi^{\rm intra}$ neglected (${\rm Im}[[M^{\rm inter}]^{-1}(\bm q,\omega)]_{11}$).
(a1), (a2) BEC regime $(k_{\rm F}a_s)^{-1}=+1$. (b1), (b2) BCS regime $(k_{\rm F}a_s)^{-1}=-1$. The decay rate is set to $\gamma_{\rm v}/\gamma_{\rm b}=0.1$ in both the figures, which are both in the unstable region. Circles in (a1) and (b1) point at poles of ${\rm det}M(\bm q,\omega)$.}
\label{fig_spcM}
\end{figure*}

Figure \ref{fig_spcM} demonstrates how the contribution from quasi-particle scatterings $\Pi^{\rm intra}$ is crucial for the appearance of dynamical instabilities.
Figures \ref{fig_spcM}(a1) and (a2) shows a spectral weight function ${\rm Im}[M^{-1}(\bm q,\omega)]_{11}$ (in the amplitude-amplitude fluctuations channel) in the BEC ($(k_{\rm F}a_s)^{-1}=+1$) and the BCS regime ($(k_{\rm F}a_s)^{-1}=-1$), respectively, where the peak structure characterizes the collective behavior of the steady-state.
In both figures, diverging spectra at the pole ${\rm det}M(\tilde{\bm q},\tilde \omega)=0$ are seen (circled), indicating that the condensate is dynamically unstable.
However, when we neglect the intra-band contribution $\Pi^{\rm intra}$ from $M(\bm q,\omega)$ by replacing $\Pi(\bm q,\omega)$ in Eq. (\ref{M}) to $\Pi^{\rm inter}(\bm q,\omega)$, i.e., 
\begin{eqnarray}
M^{\rm inter}(\bm q,\omega)
=
\left(
\begin{array}{cc}
\frac{2}{U}-\Pi_{11}^{\rm inter}(\bm q,\omega) &               -\Pi_{12}^{\rm inter}(\bm q,\omega)  \\
              -\Pi_{21}^{\rm inter}(\bm q,\omega) & \frac{2}{U}-\Pi_{22}^{\rm inter}(\bm q,\omega)
\end{array}
\right),
\nonumber\\
\label{Minter}
\end{eqnarray}
the pole disappears, as seen in Figs. \ref{fig_spcM}(a2) and (b2). 
This clearly indicates that the quasi-particle scattering is the process that triggers the dynamical instability.
Since this process can only occur when dissociated pairs are present in the steady-state, we conclude that all the dynamical instabilities we found in this paper are attributed to the pair-dissociation induced by non-equilibrium effects.

It is worth mentioning that, in Fig. \ref{fig_spcM}(b1) in the BCS side ($(k_{\rm F}a_s)^{-1}=-1$), we see a large spectral weight from the amplitude mode (also called the Higgs mode)\cite{Littlewood1981,Littlewood1982,Varma2002,Pekker2015} at $\omega\simeq 2\Delta_0 = 0.29\varepsilon_{\rm F}$, which exists in the (equilibrium) BCS superconductors\cite{Matsunaga2013}.
In contrast to the NG mode, this mode is essentially unaffected by non-equilibrium effects, since quasi-particle scattering occurs only at low energy $\omega\ll \Delta_0$, as seen in Fig. \ref{fig_spcPI}(a2).

We now discuss why the dynamical instability also occurs at non-zero $|\bm q|$ (``B'' and ``C'') in the BCS side ($(k_{\rm F}a_s)^{-1}\lesssim 0$.
In this regime, Figure \ref{fig_phase_inset}(b) shows that \textit{both} the coefficient $\alpha_{11}$ and the coefficient $\beta_{22}$ switch their sign to negative.
In this case, a ``supercurrent'' $\bm J$ (Eq. (\ref{supercurrent}))\cite{NOTEcontinuity}  anomalously flows in the \textit{anti}-parallel direction to the phase gradient $\nabla \delta\theta(\bm r,t)$.
In the BCS regime, one finds,
\begin{eqnarray}
\beta_{22}=\frac{v^2_{\mu}}{12\Delta_0^2}\alpha_{11},
\label{beta22}
\end{eqnarray}
in the small linewidth limit $\gamma\rightarrow 0^+$, where $v_{\mu}=\sqrt{2\mu/m}$. 
(For the derivation, see Appendix \ref{Appbeta22}.) 
This means that when $\alpha_{11}<0$, the anomalous supercurrent also occurs ($\beta_{22}<0$). 
When this happens, Eq. (\ref{mode}) indicates that ${\rm Im}[\omega_{\bm q}]$ becomes positive at momentum $|\bm q|>\sqrt{|\alpha_{11}|/\beta_{11}}$, leading to types ``B'' and ``C'' dynamical instabilities.
Equivalently, Eq. (\ref{detM_gam0}) qualitatively explains the non-monotonic behavior of ${\rm det}M(\bm q,\omega=0)$ found in Fig. \ref{fig_detM}(b2), recalling that $\alpha_{11}\beta_{22}\bm q^2>0$ and $\beta_{11}\beta_{22}\bm q^4<0$.
Note that higher order terms $O(\bm q^6)$ are required to explain the up-turning behavior of ${\rm det}M(\bm q,\omega=0)$ seen in Fig. \ref{fig_detM}(b2).

These ``finite-momentum-type'' instabilities may be understood from the following compensation mechanism. 
Suppose a local enhancement of the condensate fraction $\delta n_0(\bm r,t)$ occurs.
In the ordinary case when $\alpha_{11}>0$ and $\beta_{22}>0$, as described schematically in Fig. \ref{fig_instability}(a), this local enhancement decays as the condensate twists its phase $\delta\theta(\bm r,t)$ away from this spot to make a supercurrent $\bm J\sim \beta_{22}n_0\nabla\delta \theta(\bm r,t)$ flow away.
On the other hand, when the response to the amplitude fluctuations is anomalous [$\alpha_{11}<0$], for relatively small spatial modulation (low enough $|\bm q|$ that $\alpha_{11}+\beta_{11}\bm q^2<0$), the condensate twists its phase $\delta\theta(\bm r,t)$ \textit{towards} this spot (Eq. (\ref{tGL})). 
If $\beta_{22}>0$ (as in Fig. \ref{fig_instability}(b)), following Eq. (\ref{continuity}), this phase twist would make the supercurrent flow to amplify the condensate density, leading to the type ``A'' instability.
However, as illustrated in Fig. \ref{fig_instability}(d), because $\beta_{22}$ is actually negative in this regime, the supercurrent flows \textit{anti}-parallel to $\nabla \delta\theta(\bm r,t)$, stabilizing the system.
When spatial modulation is large enough (large $|\bm q|$) that $(\alpha_{11}-\beta_{11}\nabla^2)\delta n_0(\bm r,t)>0$, however, the above compensation mechanism of the supercurrent no longer works, causing the anomaly in the phase channel [$\beta_{22}<0$] to destabilize the system (as illustrated in Fig. \ref{fig_instability}(c)).
As a result, the dynamical instability starts at non-zero $|\bm q|$.

We briefly note that, although the approximate EOMs (\ref{tGL}) and (\ref{continuity}), as well as the approximate mode dispersion (\ref{mode}), are useful to understand the physics of the ``finite-momentum-type'' instabilities (as discussed above), Figure \ref{fig_phase_inset}(c) shows that the dynamical instabilities ``B'' and ``C'' occur at the decay rate slightly larger than that of where the sign change of $\alpha_{11}$ and $\beta_{22}$ occurs, while according to the approximate mode (\ref{mode}), the steady-state immediately becomes unstable at the sign change.
This is due to the neglect of amplitude-phase coupling $\alpha_{21},\beta_{12}$, etc. in Eq. (\ref{mode}), when taking the $\gamma\rightarrow 0^+$ limit.
Since this amplitude-phase coupling is associated with the violation of number conservation\cite{Watabe2012} due to the tunneling of particles between the system and the reservoirs, it works as a damping of fluctuations.
This helps the condensate to stabilize, making it possible to have regimes with both $\alpha_{11}$ and $\beta_{22}$ negative for finite $\gamma$.

\par
We finally note that a gapped mode appears in the right side of the dashed-dotted line in Fig. \ref{fig_phase}\cite{NOTE}. 
This mode may be understood as a relaxation oscillation between the condensate and the dissociated carriers\cite{Byrnes2012}. We briefly note that such a gapped mode is also seen in a driven-dissipative exciton-polariton condensate\cite{Giorgi2014}. 
\par


\section{Summary}\label{sec_summary}

To summarize, we have discussed the stability of a driven-dissipative Bose-condensed electron-hole gas in the BCS-BEC-crossover region. Within the framework of the combined BCS-Leggett-Keldysh formalism with GRPA, we showed how non-equilibrium effects lead to dynamical instabilities of an exciton-BEC. 
We find different types of dynamical instabilities, where one is due to an attractive interaction between excitons, and the other is triggered by an anomalous current that flows anti-parallel to the phase gradient of the exciton-BEC.
Our results indicate that the simple exciton-BEC characterized by the ordinary BCS-type order parameter is actually unstable in a wide region of the phase diagram in terms of the interaction strength and the decay rate. 

All of the dynamical instabilities discussed in this paper stem from pair-dissociation, induced by the non-equilibrium nature. 
Since similar depairing effects are present in an exciton-polariton condensate\cite{Yamaguchi2012}, it can be shown that dynamical instabilities arise in this non-equilibrium system as well\cite{Hanai_preparation}.
In addition, depairing effects are also present in other non-equilibrium condensates in the BCS-BEC crossover regime, such as quenched Fermi condensates in an ultracold Fermi gas\cite{Yuzbashyan2015}, as well as the neutron stars in its cooling process\cite{Yakovlev2005}, making us expect that analogous phenomena arise in these systems as well.

We end up by listing some important future problems. 
First, we have not discussed detailed physics in the shaded region in Fig. \ref{fig_phase}. 
Since this region is not necessarily in the normal phase, it is an interesting challenge to clarify what happens there. 
Inclusion of strong-coupling effects beyond the mean-field theory\cite{Kwong2009,Yoshioka2012,Nozieres1982,Hanai2013,Hanai2014} is also an important task, which would be particularly important at finite temperatures cases.
In addition, inclusions of realistic situations, such as the long-range Coulomb interaction\cite{Kwong2009,Yoshioka2012}, spin degrees of freedom\cite{Nozieres1982}, mass difference between electrons and holes\cite{Hanai2013,Hanai2014}, as well as detailed treatment of electron-hole annihilation process, would also be important for detailed comparison to experiments.

We thank M. Yamaguchi, D. Inotani, H. Tajima, and A. Edelman for discussions. This work was supported by KiPAS project in Keio University. RH was supported by a Grand-in-Aid for JSPS fellows. YO was supported by Grant-in-Aid for Scientific Research from MEXT and JSPS in Japan (No. JP15K00178, No. JP15H00840, No. JP16K05503). Work at Argonne National Laboratory is supported by the U. S. Department of Energy, BES-MSE under Contract No. DE-AC02-06CH11357.


\begin{appendix}

\section{Non-equilibrium BCS-Leggett theory}\label{AppBCSLeggett}

In our non-equilibrium BCS-Leggett theory, the self-energy is given by $\hat \Sigma=\hat \Sigma^{\rm HFB}+\hat\Sigma^{\rm env}$, diagramatically shown in Fig. \ref{figdiagram}.
\begin{widetext}
The Hartree-Fock-Bogoliubov self-energy depicted in Fig. \ref{figdiagram}(a)\cite{Hanai2016,Schrieffer} can be explicitly written as,
\begin{eqnarray}
&&\Sigma^{\rm HFB}_{\alpha,\alpha'}(\bm r,t;\bm r',t')
=iU
\sum_{\beta,\beta'=a,b}
\sum_{s,s'=\pm}\eta^{\alpha,\beta}_{\alpha',\beta'}\delta_{s,-s'}
{\rm Tr}[\tau_s {\mathscr G}_{\beta',\beta}(\bm r,t;\bm r',t')]\tau_{s'}
\delta(\bm r-\bm r')\delta(t-t'+0^+)
\nonumber\\
&&=
iU\sum_{s=\pm}\frac{1}{2}
\left(
\begin{array}{cc}
{\rm Tr}[\tau_s {\mathscr G}^{\rm K}(\bm r,t;\bm r,t+0^+)] & 0 \\
0 & {\rm Tr}[\tau_s {\mathscr G}^{\rm K}(\bm r,t;\bm r,t+0^+)]
\end{array}
\right)_{\alpha,\alpha'}
\tau_{-s}
\delta(\bm r-\bm r')\delta(t-t'+0^+)
\\
&&=
i\frac{U}{2}\sum_{s=1,2}\frac{1}{2}
\left(
\begin{array}{cc}
{\rm Tr}[\tau_s {\mathscr G}^{\rm K}(\bm r,t;\bm r,t+0^+)] & 0 \\
0 & {\rm Tr}[\tau_s {\mathscr G}^{\rm K}(\bm r,t;\bm r,t+0^+)]
\end{array}
\right)_{\alpha,\alpha'}
\tau_{s}
\delta(\bm r-\bm r')\delta(t-t'+0^+),
\label{SigHFB}
\end{eqnarray}
where $\eta^{\alpha,\alpha'}_{\beta',\beta'}=(\delta_{\alpha,\alpha'}\delta_{\beta,-\beta'}+\delta_{\alpha,-\alpha'}\delta_{\beta,\beta'})/2$.
In the third line, we transformed $\tau_{s=\pm}$ into $\tau_{s=1,2}$ for latter convenience.
The retarded component can also be written as
\begin{eqnarray}
\big[\Sigma^{\rm HFB}(\bm r,t;\bm r',t')\big]^{\rm R}
&=&-\big[
\Delta   (\bm r,t)\tau_+ 
+\Delta^*(\bm r,t)\tau_-
\big]
\delta(\bm r-\bm r')\delta(t-t'+0^+)
\\
&=&-\big[
 {\rm Re}[\Delta(\bm r,t)]\tau_1 
-{\rm Im}[\Delta(\bm r,t)]\tau_2
\big]
\delta(\bm r-\bm r')\delta(t-t'+0^+),
\label{SigHFBDel}
\end{eqnarray}
which can be derived from the definition of the order parameter,
\begin{eqnarray}
\Delta(\bm r,t)
&=&
U\avg{a_{\rm h}(\bm r,t)a_{\rm e}(\bm r,t)}
=
-i\frac{U}{2}
i\big[
\avg{a_{\rm h}(\bm r,t)a_{\rm e}(\bm r,t)}-\avg{a_{\rm e}(\bm r,t)a_{\rm h}(\bm r,t)}
\big]
=
-i\frac{U}{2}{\mathscr G}^{\rm K}_{12}(\bm r,t;\bm r,t+0^+).
\end{eqnarray}

The explicit form of Fig. \ref{figdiagram}(b) is given by,
\begin{eqnarray}
\Sigma^{\rm env}_{\alpha,\alpha'}(\bm r,t;\bm r',t') 
&=&
\sum_{\lambda={\rm b,v}}
N_{\rm t}|\Gamma_\lambda|^2
\int d\bm R d\bm R' D^\lambda_{\alpha,\alpha'}(\bm R-\bm R',t-t'),
\label{Sigenv_1}
\end{eqnarray}
describing the coupling to the bath and the vacuum within the second-order Born approximation, driving the system to a non-equilibrium state.
In obtaining Eq. (\ref{Sigenv_1}), we have taken the random average over the tunneling points $\bm r_i$ and $\bm R_i$.
Here, $D_{\alpha,\alpha'}^{\lambda={\rm b (v)}}(\bm R-\bm R',t-t')$ is the non-interacting Keldysh Green's function for the particles in the bath (vacuum), where the Fourier transformed form is given by,
\begin{eqnarray}
\hat D^\lambda(\bm q,\omega)
&=&
\left(
\begin{array}{cc}
D^\lambda_{aa} & D^\lambda_{ab} \\
D^\lambda_{ba} & D^\lambda_{bb}
\end{array}
\right)
(\bm q,\omega)
=
\left(
\begin{array}{cc}
[D^\lambda]^{\rm R} & [D^\lambda]^{\rm K} \\
0                          & [D^\lambda]^{\rm A}
\end{array}
\right)
(\bm q,\omega)
=
\left(
\begin{array}{cc}
[\omega+i\delta -\varepsilon^\lambda_{\bm q}\tau_3]^{-1} & -\pi i \tau_3 (1-2f_\lambda(\omega\tau_3)) \delta(\omega-\varepsilon^\lambda_{\bm q}\tau_3) \\
0                  & [\omega-i\delta -\varepsilon^\lambda_{\bm q}\tau_3]^{-1}
\end{array}
\right),
\nonumber\\
\end{eqnarray}
\end{widetext}
and $N_{\rm t}$ is the number of tunneling positions.
Since Eq. (\ref{Sigenv_1}) only depends on $t-t'$, we can employ Fourier transform as,
\begin{eqnarray}
\Sigma^{\rm env}_{\alpha,\alpha'}(\omega) 
=\sum_{\bm q}\sum_{\lambda={\rm b,v}}
N_{\rm t}|\Gamma_\lambda|^2
D^\lambda_{\alpha,\alpha'}(\bm q,\omega).
\end{eqnarray}
This can further be calculated by summing over $\bm q$ as, 
\begin{eqnarray}
\Sigma^{\rm env}_{\alpha,\alpha'}(\omega)
=
\sum_{\lambda={\rm b,v}}
\left(
\begin{array}{cc}
-i\gamma_\lambda & 
-2i\tau_3\gamma_\lambda [1-2f_\lambda(\omega\tau_3)]
\\
0 & i\gamma_\lambda
\end{array}
\right)_{\alpha,\alpha'}.
\label{Sigenv_2}
\end{eqnarray}
Here, $\gamma_\lambda=\pi N_{\rm t} \rho_\lambda |\Gamma_\lambda|^2$ describes the pumping ($\lambda={\rm b}$) and decay ($\lambda={\rm v}$) rate of particles. 
We have further assumed in obtaining Eq. (\ref{Sigenv_2}) that the bath and the vacuum are white, i.e., the density of states of the bath and vacuum $\rho_\lambda$ does not depend on energy $\omega$.

\section{Equation of motion with respect to $\delta\Delta(\bm r,t)$}\label{AppEOM}

We now turn to derive the equation of motion (\ref{EOM}) for a small deviation of the order parameter from the steady-state value $\delta\Delta(\bm r,t)=\Delta(\bm r,t)-\Delta_0=\delta|\Delta(\bm r,t)|+i\Delta_0\delta\theta(\bm r,t)$.
In this regard, we introduce the so-called Wigner representation\cite{Rammer}, where we transform the coordinates $(\bm r,t;\bm r',t')$ into a relative coordinate $\bm r_{\rm r}\equiv \bm r-\bm r',t_{\rm r}\equiv t-t'$ and the center of motion coordinate
$\bm r_{\rm g}=(\bm r+\bm r')/2, t_{\rm g}=(t+t')/2$.
With this representation, the retarded component of the self-energy (\ref{SigHFB}), (\ref{SigHFBDel}), (\ref{Sigenv_2}) is written as 
\begin{eqnarray}
&&
\Sigma^{\rm R}(\bm r_{\rm r},t_{\rm r};\bm r_{\rm g},t_{\rm g})
=i\frac{U}{2}\sum_{s=1,2}\frac{1}{2}
\nonumber\\
&&\times
{\rm Tr}[\tau_s {\mathscr G}^{\rm K}(\bm r_{\rm r},t_{\rm r};\bm r_{\rm g},t_{\rm g})]\tau_{s}
\delta(\bm r_{\rm r})\delta(t_{\rm r}+0^+)
-i\gamma
\nonumber\\
&&=
i\frac{U}{2}\sum_{s=1,2}\sum_{\bm p'}\int_{-\infty}^{\infty}\frac{d\omega'}{2\pi}\frac{1}{2}{\rm Tr}[\tau_s {\mathscr G}^{\rm K}(\bm p',\omega';\bm r_{\rm g},t_{\rm g})]\tau_{s} 
-i\gamma
\nonumber\\
\\
&&=-\big[
{\rm Re}[\Delta(\bm r_{\rm g},t_{\rm g})]\tau_1-{\rm Im}[\Delta(\bm r_{\rm g},t_{\rm g})]\tau_2
\big]
-i\gamma.
\end{eqnarray}
Thus, the small deviation of the self-energy from the steady-state value $\delta\Sigma^{\rm R}=\Sigma^{\rm R}-\Sigma^{\rm R}_0$ is given by,
\begin{eqnarray}
\delta\Sigma^{\rm R}(\bm p,\nu;\bm r_{\rm g},t_{\rm g})
&=&
i\frac{U}{2}\sum_{s=1,2}\sum_{\bm p'}\int_{-\infty}^{\infty}\frac{d\omega'}{2\pi}
\frac{1}{2}
\nonumber\\
&\times&
{\rm Tr}[\tau_s\delta G^{\rm K}(\bm p',\omega';\bm r_{\rm g},t_{\rm g})]\tau_s
\label{deltaSig}
\\
&=&
-\big[
\delta|\Delta(\bm r_{\rm g},t_{\rm g})|\tau_1 - \Delta_0\delta\theta(\bm r_{\rm g},t_{\rm g})\tau_2
\label{deltaSigDel}
\big],
\nonumber\\
\end{eqnarray}
where $\delta G^{\rm K}(\bm p,\nu;\bm r_{\rm g},t_{\rm g})={\mathscr G}^{\rm K}(\bm p,\nu;\bm r_{\rm g},t_{\rm g})-G^{\rm K}(\bm p,\nu)$ is a small deviation of the Keldysh component of the single-particle Green's function from the steady-state, driven by $\delta\Delta(\bm r_{\rm g},t_{\rm g})$.

To proceed, it would be convenient to introduce a short-handed notation\cite{Rammer},
\begin{eqnarray}
(A\otimes B)(\bm r,t;\bm r',t')
&=&
\int d\bm r_1 \int_{-\infty}^{\infty}dt_1
\nonumber\\
&\times &
A(\bm r,t;\bm r_1,t_1)B(\bm r_1,t_1;\bm r',t').
\end{eqnarray}
With this notation, the Dyson's equation (\ref{Dyson}) can be represented as,
\begin{eqnarray}
\hat {\mathscr G}=\hat G^{0}+\hat G^{0}\otimes \hat \Sigma \otimes \hat {\mathscr G},
\end{eqnarray}
or
\begin{eqnarray}
([\hat G^0]^{-1} - \hat \Sigma)\otimes \hat {\mathscr G}= {\bm 1}, 
\label{Dysonshort}
\end{eqnarray}
where
\begin{eqnarray}
&&([\hat G^0]^{-1} \otimes \hat{\mathscr G})(\bm r,t;\bm r',t')
\nonumber\\
&&=
\left (
\begin{array}{cc}
i\partial_t - (- \frac{\nabla^2}{2m} - \mu) \tau_3 & 0 \\
0 & i\partial_t - (-\frac{\nabla^2}{2m} - \mu)\tau_3
\end{array}
\right )
\nonumber\\
&&\times
\hat{\mathscr G}(\bm r,t;\bm r',t').
\end{eqnarray}
From Eq. (\ref{Dysonshort}), the retarded component ${\mathscr G}^{\rm R}$ is given by,
\begin{eqnarray}
([G^{0{\rm R}}]^{-1} - \Sigma^{\rm R})\otimes {\mathscr G}^{\rm R}= {\bm 1}, 
\end{eqnarray}
which yields,
\begin{eqnarray}
-\delta\Sigma^{\rm R}\otimes G^{\rm R} +[G^{\rm R}]^{-1} \otimes \delta G^{\rm R}=0,
\end{eqnarray}
where $\delta G^{\rm R}={\mathscr G}^{\rm R}-G^{\rm R}$.  
Thus, we obtain  
\begin{eqnarray}
\delta G^{\rm R}
&=&
 G^{\rm R}\otimes \delta \Sigma^{\rm R}\otimes G^{\rm R}
\nonumber\\ 
&=&
-G^{\rm R}\otimes [\delta|\Delta|\tau_1 - \Delta_0\delta\theta\tau_2 ]\otimes G^{\rm R},
\label{deltaGR}
\end{eqnarray}
where we have used Eq. (\ref{deltaSigDel}) in the last equality.
The advanced component $\delta G^{\rm A}={\mathscr G}^{\rm A}-G^{\rm A}$ is immediately obtained by taking the Hermite conjugate of the retarded component, i.e., 
\begin{eqnarray}
\delta G^{\rm A}=[\delta G^{\rm R}]^\dagger
=-G^{\rm A}\otimes [\delta|\Delta|\tau_1 - \Delta_0\delta\theta\tau_2 ]\otimes G^{\rm A}.
\nonumber\\
\label{deltaGA}
\end{eqnarray}

The Keldysh component of the single-particle Green's function ${\mathscr G}^{\rm K}$ can be obtained from the Dyson's equation (\ref{Dysonshort}) as,
\begin{eqnarray}
[G^{0{\rm R}}]^{-1}\otimes {\mathscr G}^{\rm K}
-\Sigma^{\rm R}\otimes {\mathscr G}^{\rm K}-\Sigma^{\rm K}\otimes {\mathscr G}^{\rm A}=0,
\end{eqnarray}
which relates $\delta G^{\rm K}$ to $\delta G^{\rm R/A}$ as, (Note that $\Sigma^{\rm K}=\Sigma^{\rm K}_0$ within our approximation.)
\begin{eqnarray}
[G^{\rm R}]^{-1}\otimes \delta G^{\rm K}
-\delta\Sigma^{\rm R}\otimes G^{\rm K}
-\Sigma^{\rm K}_0 \otimes \delta G^{\rm A} =0,
\end{eqnarray}
or 
\begin{widetext}
\begin{eqnarray}
\delta G^{\rm K}
&=&
G^{\rm R}\otimes \delta \Sigma^{\rm R} \otimes G^{\rm K}+G^{\rm R}\otimes\Sigma^{\rm K}_0\otimes \delta G^{\rm A}
\nonumber\\
&=&-\Big[
G^{\rm R}\otimes [\delta|\Delta|\tau_1 - \Delta_0\delta\theta\tau_2 ]\otimes G^{\rm K}
+G^{\rm R}\otimes\Sigma^{\rm K}_0 \otimes G^{\rm A}\otimes[\delta|\Delta|\tau_1 - \Delta_0\delta\theta\tau_2 ] \otimes G^{\rm A}
\Big]
\nonumber\\
&=&-
\Big[
G^{\rm R}\otimes [\delta|\Delta|\tau_1 - \Delta_0\delta\theta\tau_2 ]\otimes G^{\rm K}
+G^{\rm K}\otimes[\delta|\Delta|\tau_1 - \Delta_0\delta\theta\tau_2 ] \otimes G^{\rm A}
\Big].
\end{eqnarray}
Here, we have used Eqs. (\ref{deltaSigDel}), (\ref{deltaGA}) 
in obtaining the second line, and Eq. (\ref{Gk}) in obtaining the third.

When $\delta\Delta(\bm r_{\rm g},t_{\rm g})= \delta\Delta(\bm q,\omega) e^{i\bm q\cdot \bm r_{\rm g}-i\omega t_{\rm g}}$, 
\begin{eqnarray}
&&\delta G^{\rm K}(\bm p,\nu;\bm r_{\rm g},t_{\rm g})
=-\int d\bm r_{\rm r} \int_{-\infty}^{\infty} dt_{\rm r} e^{-i\bm p\cdot \bm r_{\rm r} + i\nu t_{\rm r}}\int d\bm r_1 \int_{-\infty}^{\infty} dt_1 
\nonumber\\
&&\ \ \ \ \times 
\bigg[
G^{\rm R}
\Big(
\bm r_{\rm g}+\frac{\bm r_{\rm r}}{2}-\bm r_1,t_{\rm g}+\frac{t_{\rm r}}{2}-t_1
\Big)
\Big[
\delta|\Delta(\bm r_1,t_1)|\tau_1 - \Delta_0 \delta\theta(\bm r_1,t_1)\tau_2
\Big]
G^{\rm K}
\Big(
-\bm r_{\rm g}+\frac{\bm r_{\rm r}}{2}+\bm r_1,-t_{\rm g}+\frac{t_{\rm r}}{2}+t_1
\Big)
\nonumber\\
&&\ \ \ \ \ \ +
G^{\rm K}
\Big(
\bm r_{\rm g}+\frac{\bm r_{\rm r}}{2}-\bm r_1,t_{\rm g}+\frac{t_{\rm r}}{2}-t_1
\Big)
\Big[
\delta|\Delta(\bm r_1,t_1)|\tau_1 - \Delta_0 \delta\theta(\bm r_1,t_1)\tau_2
\Big]
G^{\rm A}
\Big(
-\bm r_{\rm g}+\frac{\bm r_{\rm r}}{2}+\bm r_1,-t_{\rm g}+\frac{t_{\rm r}}{2}+t_1
\Big)
\bigg ]
\nonumber\\
&& \ \ =
-\int d\bm r_{\rm r} \int_{-\infty}^{\infty} dt_{\rm r} e^{-i\bm p\cdot \bm r_{\rm r} + i\nu t_{\rm r}}\int d\bm r_1 \int_{-\infty}^{\infty} dt_1 
\nonumber\\
&&\ \ \times 
\sum_{\bm p'}\sum_{\bm p''}\sum_{\bm q'}
\int \frac{d\nu '}{2\pi} \frac{d\nu''}{2\pi} \frac{d\omega'}{2\pi}
e^{i\bm p'\cdot (\bm r_{\rm g}+\bm r_{\rm r}/2-\bm r_1) - i\nu' (t_{\rm g}+t_{\rm r}/2-t_1)}
e^{i\bm q'\cdot \bm r_1-i\omega' t_1}
e^{i\bm p''\cdot (-\bm r_{\rm g}+\bm r_{\rm r}/2+\bm r_1) - i\nu'' (-t_{\rm g}+t_{\rm r}/2+t_1)}
\nonumber\\
&&\ \ \ \ \ \ \ \ \ \ \ \ \ \ \ \ \ \ 
\times
\bigg[
G^{\rm R} ( \bm p',\nu' )
\Big[
\delta|\Delta(\bm q',\omega')|\tau_1 - \Delta_0\delta\theta(\bm q',\omega')\tau_2
\Big]
\delta_{\bm q,\bm q'}
2\pi \delta(\omega-\omega')
G^{\rm K}( \bm p'',\nu'' )
\nonumber\\
&&\ \ \ \ \ \ \ \ \ \ \ \ \ \ \ \ \ \ \ \ 
+
G^{\rm K} ( \bm p',\nu' )
\Big[
\delta|\Delta(\bm q',\omega')|\tau_1 - \Delta_0\delta\theta(\bm q',\omega')\tau_2
\Big]
\delta_{\bm q,\bm q'}
2\pi \delta(\omega-\omega')
G^{\rm A}( \bm p'',\nu'' )
\bigg]
\nonumber\\
&& \ \ =
- e^{i\bm q\cdot \bm r_{\rm g} - i\omega t_{\rm g}}
\bigg[
G^{\rm R} \Big( \bm p+\frac{\bm q}{2},\nu+\frac{\omega}{2}\Big)
\Big[
\delta|\Delta(\bm q,\omega)|\tau_1 - \Delta_0\delta\theta(\bm q,\omega)\tau_2
\Big]
G^{\rm K}\Big( \bm p - \frac{\bm q}{2},\nu-\frac{\omega}{2} \Big)
\nonumber\\
&&\ \ \ \ \ \ \ \ \ \ \ \ \ \ \ \ \ \ \ \ 
+
G^{\rm K} \Big( \bm p+\frac{\bm q}{2},\nu+\frac{\omega}{2}\Big)
\Big[
\delta|\Delta(\bm q,\omega)|\tau_1 - \Delta_0\delta\theta(\bm q,\omega)\tau_2
\Big]
G^{\rm A}\Big( \bm p - \frac{\bm q}{2},\nu-\frac{\omega}{2} \Big)
\bigg].
\label{deltaGk}
\end{eqnarray}
Substituting Eq. (\ref{deltaGk}) into Eq. (\ref{deltaSigDel}) and equating it with Eq. (\ref{deltaSig}), we obtain,
\begin{eqnarray}
&&\delta|\Delta(\bm q,\omega)|\tau_1 - \Delta_0\delta\theta(\bm q,\omega)\tau_2
\nonumber\\
&& 
=
\frac{U}{2}\sum_{s=1,2}
\frac{i}{2}
\sum_{\bm p}\int_{-\infty}^{\infty}\frac{d\nu}{2\pi}
{\rm Tr} 
\bigg[
\tau_s
G^{\rm R} \Big( \bm p+\frac{\bm q}{2},\nu+\frac{\omega}{2}\Big)
\Big[
\delta|\Delta(\bm q,\omega)|\tau_1 - \Delta_0\delta\theta(\bm q,\omega)\tau_2
\Big]
G^{\rm K}\Big( \bm p - \frac{\bm q}{2},\nu-\frac{\omega}{2}\Big)
\nonumber\\
&&\ \ \ \ \ \ \ \ \ \ \ \ \ \ \ \ \ \ \ \ \ \ \ \ \ \ \ \ \ \ \ 
+
\tau_s
G^{\rm K} \Big( \bm p+\frac{\bm q}{2},\nu+\frac{\omega}{2}\Big)
\Big[
\delta|\Delta(\bm q,\omega)|\tau_1 - \Delta_0\delta\theta(\bm q,\omega)\tau_2
\Big]
G^{\rm A}\Big( \bm p - \frac{\bm q}{2},\nu-\frac{\omega}{2} \Big)
\bigg]\tau_s,
\end{eqnarray}
or
\begin{eqnarray}
\left(
\begin{array}{c}
\delta|\Delta(\bm q,\omega)|  \\
-\Delta_0\delta\theta(\bm q,\omega) 
\end{array}
\right)
=
\frac{U}{2}
\left(
\begin{array}{cc}
\Pi_{11}(\bm q,\omega) & \Pi_{12}(\bm q,\omega)  \\
\Pi_{21}(\bm q,\omega) & \Pi_{22}(\bm q,\omega)
\end{array}
\right)
\left(
\begin{array}{c}
\delta|\Delta(\bm q,\omega)|  \\
-\Delta_0\delta\theta(\bm q,\omega) 
\end{array}
\right),
\end{eqnarray}
where
\begin{eqnarray}
&&\Pi_{s,s'}(\bm q,\omega)
=
\frac{i}{2}
\sum_{\bm p}\int_{-\infty}^{\infty}\frac{d\nu}{2\pi}
{\rm Tr} 
\bigg[
\Big[
\tau_s
\hat G\Big( \bm p+\frac{\bm q}{2},\nu+\frac{\omega}{2}\Big)
\tau_{s'}
\hat G\Big( \bm p - \frac{\bm q}{2},\nu-\frac{\omega}{2} \Big)
\Big]^{\rm K}
\bigg].
\end{eqnarray}
We have used the short-hand notation introduced in Eqs. (\ref{shortR})-(\ref{shortK}).
Dividing both sides by $U/2$, and transposing the right-hand side to the left-hand side gives the desired Eq. (\ref{EOM}).
\end{widetext}

\section{Lower and upper branch contributions to single-particle Green's function}\label{App_Gul}

We give here the definition of the lower ($\hat G_{\rm l}$) and the upper ($\hat G_{\rm u}$) branch contribution of the single-particle Green's function $\hat G(=\hat G_{\rm l}+\hat G_{\rm u})$.
First, the retarded component (\ref{Gra}) can be split into the lower and the upper branch contribution as
\begin{eqnarray}
G^{\rm R}(\bm p,\omega)&=&G^{\rm R}_{\rm u}(\bm p,\omega)+G^{\rm R}_{\rm l}(\bm p,\omega),
\label{grga}
\end{eqnarray}
where
\begin{eqnarray}
G^{\rm R}_{\rm l}(\bm p,\omega)
&=&
\frac{1}{2}\Big[\bm 1 - \frac{\xi_{\bm p}}{E_{\bm p}}\tau_3- \frac{\Delta_0}{E_{\bm p}} \tau_1
\Big]
\frac{1}{\omega + i\gamma +E_{\bm p}}.
\label{gral}\\
G^{\rm R}_{\rm u}(\bm p,\omega)
&=&
\frac{1}{2}\Big[\bm 1 + \frac{\xi_{\bm p}}{E_{\bm p}}\tau_3 + \frac{\Delta_0}{E_{\bm p}} \tau_1
\Big]
\frac{1}{\omega + i\gamma - E_{\bm p}}.
\label{grau}
\end{eqnarray}
Since $G^{\rm R}_{\rm l}(\bm p,\omega)$ and $G^{\rm R}_{\rm u}(\bm p,\omega)$ give the dispersion at $\omega=-E_{\bm p}$ and $\omega=E_{\bm p}$ with linewidth $\gamma$, respectively, these correspond to the lower and the upper branch contribution, respectively. The advanced component of the lower ($G_{\rm l}^{\rm A}$) and the upper branch ($G_{\rm u}^{\rm A}$) are given respectively by,
\begin{eqnarray}
G^{\rm A}_{\rm l}(\bm p,\omega) &=& [G^{\rm R}_{\rm l}(\bm p,\omega)]^\dagger, \\
G^{\rm A}_{\rm u}(\bm p,\omega) &=& [G^{\rm R}_{\rm u}(\bm p,\omega)]^\dagger.
\end{eqnarray}

The Keldysh component (\ref{Gk}) can also be split into the upper and lower branch contributions by use of Eq. (\ref{grga}). Since the Keldysh component is anti-Hermitian (i.e., $G^{\rm K}(\bm p,\omega)=-[G^{\rm K}(\bm p,\omega)]^\dagger$)\cite{Rammer}, it can be expressed with $G^{\rm R}(\bm p,\omega)$ and $G^{\rm A}(\bm p,\omega)$ as
\begin{eqnarray}
G^{\rm K}(\bm p,\omega)
&=&G^{\rm R}(\bm p,\omega)F(\bm p,\omega)-[G^{\rm R}(\bm p,\omega)F(\bm p,\omega)]^\dagger
\nonumber\\
&=&G^{\rm R}(\bm p,\omega)F(\bm p,\omega)-F(\bm p,\omega)G^{\rm A}(\bm p,\omega),
\end{eqnarray}
where $F(\bm p,\omega)(=F^\dagger(\bm p,\omega))$ has information on the distribution function.
The explicit form is written as
\begin{eqnarray}
F(\bm p,\omega)
&=&
F_{-}(\omega)\bm 1+\frac{\xi_{\bm p}^2+\gamma^2}{E_{\bm p}^2+\gamma^2}F_{+}(\omega)\tau_3 
\nonumber\\
&+& \frac{\Delta_0\xi_{\bm p}}{E_{\bm p}^2+\gamma^2}F_{+}(\omega)\tau_1- \frac{\Delta_0\gamma}{E_{\bm p}^2+\gamma^2}F_{+}(\omega)\tau_2.
\end{eqnarray}
Thus, by using Eq. (\ref{grga}), we can also split the Keldysh component as
\begin{eqnarray}
G^{\rm K}(\bm p,\omega)&=&G^{\rm K}_{\rm u}(\bm p,\omega)+G^{\rm K}_{\rm l}(\bm p,\omega),
\label{gk}\\
G^{\rm K}_{\rm l}(\bm p,\omega)&=&G^{\rm R}_{\rm l}(\bm p,\omega)F(\bm p,\omega)-F(\bm p,\omega)G^{\rm A}_{\rm l}(\bm p,\omega), \\
G^{\rm K}_{\rm u}(\bm p,\omega)&=&G^{\rm R}_{\rm u}(\bm p,\omega)F(\bm p,\omega)-F(\bm p,\omega)G^{\rm A}_{\rm u}(\bm p,\omega),
\nonumber\\
\end{eqnarray}
where $G_{\rm l}^{\rm K}$ and $G_{\rm u}^{\rm K}$ describe the lower and the upper branch contribution, respectively.

\section{Derivation of Eq. (\ref{beta22})}\label{Appbeta22}

Here, we derive Eq. (\ref{beta22}) in the main text that holds in the BCS regime by expanding $M_{22}(\bm q,0)$ in powers of $\bm q$.
To do so, we first expand the single-particle Green's function $G^{\rm R/A/K}(\bm k+\bm q,\omega)$, similarly to Eqs. (\ref{GrDelseries}), (\ref{GaDelseries}), and (\ref{GkDelseries}).
We find from Eq. (\ref{Gra}) that 
\begin{eqnarray}
[G^{\rm R}(\bm k+\bm q,\omega )]^{-1}
&=&(\omega + i\gamma) \bm 1 - \xi_{\bm k + \bm q}\tau_3+\Delta_0 \tau_1
\nonumber\\
&=& [G^{\rm R}(\bm k,\omega)]^{-1}-\delta \xi_{\bm k \bm q}\tau_3,
\end{eqnarray}
where $\xi_{\bm k+\bm q}=\xi_{\bm k}+\delta \xi_{\bm {kq}}=\xi_{\bm k}+\bm v_{\bm k}\cdot \bm q+{\bm q}^2/(4m)$, 
making it possible to expand $G^{\rm R}(\bm k+\bm q,\omega)$ as
\begin{eqnarray}
G^{\rm R}(\bm k +\bm q,\omega )
&=& G^{\rm R}(\bm k,\omega)
+ [\hat G(\bm k,\omega)\tau_3 \hat G(\bm k,\omega)]^{\rm R}\delta\xi_{\bm k\bm q}
\nonumber\\
&+&
[\hat G(\bm k,\omega)\tau_3 \hat G(\bm k,\omega)\tau_3 \hat G(\bm k,\omega)]^{\rm R}\delta\xi_{\bm k\bm q}^2 + \cdots.
\nonumber\\
\label{Graseries}
\end{eqnarray}
We also obtain from Eqs. (\ref{Gk_1}) and (\ref{Graseries}), 
\begin{eqnarray}
&&G^{\rm K}(\bm k+\bm q,\omega )
=G^{\rm R}(\bm k+\bm q,\omega )\Sigma^{\rm K}_0(\omega)G^{\rm A}(\bm k+\bm q,\omega )
\nonumber\\
&&
= G^{\rm K}(\bm k,\omega)
+ [\hat G(\bm k,\omega)\tau_3 \hat G(\bm k,\omega)]^{\rm K}\delta\xi_{\bm k\bm q}
\nonumber\\
&&
+[\hat G(\bm k,\omega)\tau_3 \hat G(\bm k,\omega)\tau_3 \hat G(\bm k,\omega)]^{\rm K}\delta\xi_{\bm k\bm q}^2 + \cdots,
\label{Gkseries}
\end{eqnarray}
enabling us to expand $G^{\rm K}(\bm k+\bm q,\omega)$ in terms of $\delta\xi_{\bm k\bm q}$ as well, noting $G^{\rm A}=[G^{\rm R}]^\dagger$.

We now restrict ourselves to the BCS regime. In this regime, since the dominant contribution in the $\bm k$-summention is at $|\bm k|\simeq \sqrt{2m \mu}=m v_\mu \gg |\bm q|$ ($\mu>0$), we can approximate 
\begin{eqnarray}
\delta \xi_{\bm k\bm q}\simeq \bm v_{\mu}\cdot \bm q.
\label{deltaxikq}
\end{eqnarray}
\begin{widetext}
Using Eqs. (\ref{Graseries})-(\ref{deltaxikq}), we obtain
\begin{eqnarray}
&&M_{22}(\bm q,0)
=\frac{2}{U}-\frac{i}{2}\sum_{\bm k}\int \frac{d\omega}{2\pi}
{\rm Tr} 
\big [
\tau_2 G^{\rm R}(\bm k+\bm q,\omega)\tau_2 G^{\rm K}(\bm k,\omega)
+\tau_2 G^{\rm K}(\bm k+\bm q,\omega)\tau_2 G^{\rm A}(\bm k,\omega)
\big ]
\nonumber\\
&&  
\simeq 
-\frac{i}{2}\sum_{\bm k}\int \frac{d\omega}{2\pi}
{\rm Tr} 
\Big [
\big[
\tau_2 \hat G(\bm k,\omega)\tau_3 \hat G(\bm k,\omega)\tau_3 \hat G(\bm k,\omega)\tau_2 \hat G(\bm k,\omega)
\big ]^{\rm K}
\Big ]
(\bm v_\mu\cdot \bm q)^2 + O(\bm q^4)
\nonumber\\
\nonumber\\
&&   
\simeq 
-i\rho(0) \frac{v_{\mu}^2 q^2}{3} \int_{-\infty}^{\infty} d\xi \int \frac{d\omega}{2\pi}
{\rm Tr} 
\Big [
\big [
\tau_2 \hat G(\bm k,\omega)\tau_3 \hat G(\bm k,\omega)\tau_3 \hat G(\bm k,\omega)\tau_2 \hat G(\bm k,\omega)
\big]^{\rm K}
\Big ]
+ O(\bm q^4),
\label{pi22expand}
\end{eqnarray}
where we have used the Thouless criterion (\ref{ThoulessRe}).
In the third line of Eq. (\ref{pi22expand}), we replaced the density of states $\rho(\omega)$ of free carriers with that at $\omega=0$, when replacing the $\bm k$-summention by a $\xi(\equiv k^2/(2m)-\mu)$-integral. This is justified when $\mu>0$ and $\mu \gg \gamma,\Delta_0$.
This thus gives
\begin{eqnarray}
&&\beta_{22}
=
-i\rho(0) 
\frac{v_{\mu}^2}{3}
\int_{-\infty}^{\infty} d\xi \int \frac{d\omega}{2\pi}
{\rm Tr} 
\Big [
\big[\tau_2 \hat G(\bm k,\omega)\tau_3 \hat G(\bm k,\omega)\tau_3 \hat G(\bm k,\omega)\tau_2 \hat G(\bm k,\omega)
\big ]^{\rm K}
\Big ]
\nonumber\\
&&=
-i\rho(0) 
\frac{v_{\mu}^2}{6}
\int_{-\infty}^{\infty} d\xi \int \frac{d\omega}{2\pi}
{\rm Tr} 
\Big [
\big[\tau_2 \hat G(\bm k,\omega)\tau_3 \hat G(\bm k,\omega)\tau_3 \hat G(\bm k,\omega)\tau_2 \hat G(\bm k,\omega)
\nonumber\\
&& \ \ \ \ \ \ \ \ \ \ \ \ \ \ \ \ \ \ \ \ \ \ \ \ \ \ \ \ \ \ \ \ \ \ \ \ \ 
+\tau_3 \hat G(\bm k,\omega)\tau_2 \hat G(\bm k,\omega)\tau_2 \hat G(\bm k,\omega)\tau_3 \hat G(\bm k,\omega) 
\big]^{\rm K}
\Big ]
\nonumber\\
&&=
i\rho(0) 
\frac{v_{\mu}^2}{6}
\int_{-\infty}^{\infty} d\xi 
\int \frac{d\omega}{2\pi}
\nonumber\\
&&\times
\big [
(G_{12}G_{12}-G_{22}G_{22})(-G_{12}G_{12}+G_{22}G_{11})
+(G_{11}G_{11}-G_{12}G_{12})(G_{11}G_{22}-G_{12}G_{12})
\nonumber\\
&&+(-G_{12}G_{12}+G_{22}G_{11})(G_{12}G_{12}-G_{22}G_{22})
+(G_{11}G_{22}-G_{12}G_{12})(G_{11}G_{11}-G_{12}G_{12})
\nonumber\\
&&+(G_{12}G_{11}-G_{22}G_{12})(G_{11}G_{12}-G_{12}G_{11})
+(-G_{11}G_{12}+G_{12}G_{22})(-G_{12}G_{22}+G_{22}G_{12})
\nonumber\\
&&+(G_{11}G_{12}-G_{12}G_{11})(G_{12}G_{11}-G_{22}G_{12})
+(-G_{12}G_{22}+G_{22}G_{12})(-G_{11}G_{12}+G_{12}G_{22})
\big ]^{\rm K},
\label{piqr22}
\end{eqnarray}
where we omitted $(\bm k,\omega)$ in the third line.
In obtaining the second line in Eq. (\ref{piqr22}), we have used the cyclic property of the trace and
Eq. (\ref{Grak12}) in the third.
(The Keldysh component in Eq. (\ref{Grak12}) holds only in the small linewidth limit $\gamma \rightarrow 0^+$.)

By further using Eq. (\ref{G12Del}), as well as
\begin{eqnarray}
-i\frac{G^{\rm R/A}_{12}(\bm k,\omega)}{\Delta_0} 
&=&\frac{i}{2}{\rm Tr}\Big[
\tau_2 \hat G(\bm k,\omega) \tau_2 \hat G(\bm k,\omega)
\Big]^{\rm R/A},
\end{eqnarray}
we obtain,
\begin{eqnarray}
\beta_{22}
&=&
-i\rho(0) 
\frac{v_{\mu}^2}{6}
\int_{-\infty}^{\infty} d\xi 
\int \frac{d\omega}{2\pi}
\Big[ 
4\frac{G_{12}(\bm k,\omega)}{\Delta_0}\frac{G_{12}(\bm k,\omega)}{\Delta_0}
\nonumber\\
&&\ \ \ \ \ \ \ \ \ \ \ \ \ \ \ \   
+(G_{11}(\bm k,\omega)-G_{22}(\bm k,\omega))(G_{11}(\bm k,\omega)-G_{22}(\bm k,\omega))\frac{G_{12}(\bm k,\omega)}{\Delta_0}
\nonumber\\
&&\ \ \ \ \ \ \ \ \ \ \ \ \ \ \ \ 
+\frac{G_{12}(\bm k,\omega)}{\Delta_0}(G_{11}(\bm k,\omega)-G_{22}(\bm k,\omega))(G_{11}(\bm k,\omega)-G_{22}(\bm k,\omega))
\Big]^{\rm K}.
\end{eqnarray}
Here, we have also used Eqs. (\ref{gr11momega})-(\ref{gk12momega}) to drop the antisymmetric part from the $\omega$-integral.
In fact, we can show that
\begin{eqnarray}
&&\int_{-\infty}^{\infty} d\xi
\Big[
2\frac{G_{12}(\bm k,\omega)}{\Delta_0}\frac{G_{12}(\bm k,\omega)}{\Delta_0} 
+(G_{11}(\bm k,\omega)-G_{22}(\bm k,\omega))(G_{11}(\bm k,\omega)-G_{22}(\bm k,\omega))\frac{G_{12}(\bm k,\omega)}{\Delta_0}
\nonumber\\
&& \ \ \ \ \ \ \ \ \ \ \ \ \ \ \ \ \ \ \ \ \ 
+\frac{G_{12}(\bm k,\omega)}{\Delta_0}(G_{11}(\bm k,\omega)-G_{22}(\bm k,\omega))(G_{11}(\bm k,\omega)-G_{22}(\bm k,\omega))
\Big]^{\rm K}=0,
\end{eqnarray}
which thus gives 
\begin{eqnarray}
\beta_{22}=-i\rho(0) 
\frac{v_{\mu}^2}{3\Delta_0^2}
\int_{-\infty}^{\infty} d\xi 
\int \frac{d\omega}{2\pi}
\Big[ 
G^{\rm R}_{12}(\bm k,\omega)G^{\rm K}_{12}(\bm k,\omega)+G^{\rm K}_{12}(\bm k,\omega)G^{\rm A}_{12}(\bm k,\omega)
\Big].
\label{beta22final}
\end{eqnarray}
On the other hand, $\alpha_{11}$ in the BCS regime can be expressed as 
\begin{eqnarray}
\alpha_{11}
&=&\frac{2}{U}-2i\rho(0)\int_{-\infty}^{\infty} d\xi 
\int \frac{d\omega}{2\pi}
\Big[ 
G^{\rm R}_{12}(\bm k,\omega)G^{\rm K}_{12}(\bm k,\omega)+G^{\rm K}_{12}(\bm k,\omega)G^{\rm A}_{12}(\bm k,\omega)
\Big]\nonumber\\
&=&
-4i\rho(0)\int_{-\infty}^{\infty} d\xi 
\int \frac{d\omega}{2\pi}
\Big[ 
G^{\rm R}_{12}(\bm k,\omega)G^{\rm K}_{12}(\bm k,\omega)+G^{\rm K}_{12}(\bm k,\omega)G^{\rm A}_{12}(\bm k,\omega)
\Big],
\label{piq011}
\end{eqnarray}
where we have used Eqs. (\ref{ThoulessRe}).
Combining Eqs. (\ref{beta22final}) and (\ref{piq011}) gives the desired Eq. (\ref{beta22}). 

\end{widetext}

\end{appendix}


\begin{thebibliography}{99}
\bibitem{NOTEcontinuity} We briefly note that Eq. (\ref{continuity}) is an approximate continuity equation up to $O(\partial_t,\nabla^2)$.
\bibitem{NOTEDeltaexpand} Precisely speaking, $\Delta_0\ll|\mu_{\rm b}|$ is also necessary to justify the expansion in terms of $\Delta_0$.
\bibitem{NOTE} In this paper, the gapped mode is determined from the real part of the mode equation, ${\rm Re}[{\rm det}[M(\bm q\rightarrow 0,\omega_{\bm q})]]=0$. Since ${\rm Im}[{\rm det}[M(\bm q\rightarrow 0,\omega_{\bm q})]]=0$ is not necessarily satisfied, the gapped mode may actually be diffusive.
\end{thebibliography}
\end{document}